\documentclass[aps, prb, reprint, twocolumn,floatfix]{revtex4-2}
\usepackage{graphicx} 

\usepackage{dblfloatfix}

\usepackage{dcolumn}
\usepackage{bm}
\usepackage[version=4]{mhchem}
\usepackage{siunitx}

\usepackage{hyperref}
\usepackage{xcolor}
\newcommand{\tl}[1]{{\color{cyan} TL: #1}}
\newcommand{\jws}[1]{{\color{red} JWS: #1}}
\newcommand{\jv}[1]{{\color{violet} JV: #1}}

\usepackage{comment}

\begin{document}


\title{Rationalizing defect formation energies in metals and semiconductors with semilocal density functionals}


\author{Jorge Vega Bazantes}
\author{Timo Lebeda}
\email[Contact author: tl681@cam.ac.uk; Current address: Yusuf Hamied Department of Chemistry,
University of Cambridge, Cambridge CB2 1EW, UK]{}
\author{Akilan Ramasamy}
\author{Kanun Pokharel}
\author{Ruiqi Zhang}
\author{John P. Perdew}
\author{Jianwei Sun}\thanks{Contact author: jsun@tulane.edu}

\affiliation{%
Department of Physics and Engineering Physics, Tulane University, New Orleans, Louisiana 70118, USA
}%


\begin{abstract}
The study of defects in materials is of utmost importance for technological applications and the design of new materials. In this work, we analyze the performance of density functional approximations on two prototypical sets of defective systems: monovacancies in eight fcc metals, and interstitials in the semiconductor Si-diamond. Specifically, we compute defect formation energies using the local density approximation, the Perdew–Burke–Ernzerhof generalized gradient approximation, the meta-generalized gradient approximations (meta-GGAs) strongly constrained and appropriately normed (SCAN), its regularized-restored version (r$^2$SCAN), the Lebeda--Aschebrock--K\"ummel (LAK) meta–GGA, and the Heyd--Scuseria--Ernzerhof screened hybrid functional. For metals, the local density approximation shows better performance compared to the other approximations, whereas for silicon, the LAK meta-GGA yields outstanding accuracy, surpassing the hybrid functional and approaching the results of more computationally demanding quantum Monte Carlo methods. To rationalize the different performances, we study the semilocal ingredients $r_s$, $s$, and $\alpha$ in both the pristine and defective structures. We identify critical regions that indicate the observed trends of the defect formation energies and pave the way for improving density functional approximations.

\end{abstract}

\maketitle


\section{\label{sec:level1}Introduction}

Point defects play an essential role in determining the properties of materials, influencing electronic, optical, and catalytic behavior across a wide range of applications \cite{li2020defect, pastor2022electronic, mosquera2023imperfections, tan2026multifunctional}. In particular, defect formation energies govern the stability and concentration of defects, thereby controlling key processes such as diffusion, doping, and phase transformations in both metals and semiconductors \cite{freysoldt2014first}. These effects are not only critical for conventional technologies but also underpin emerging applications such as quantum computing \cite{de2021materials, wolfowicz2020vanadium, bosma2018identification}. 

On the theoretical side, Kohn--Sham density functional theory (DFT) \cite{kohn1965self, parr1989density, perdew2003density} has become a cornerstone of \textit{ab initio} condensed matter physics over the past decades. In principle, DFT provides the exact ground-state density \(n(\mathbf{r})\) and total energy \(E\) of interacting electrons through self-consistent one-electron equations \cite{kohn1965self}. In practice, however, the exchange--correlation (xc) functional \(E_\mathrm{xc}\) must be approximated. Within the non-empirical paradigm, these approximations are constructed by satisfying exact constraints and appropriate norms \cite{sun2015strongly,kaplan2023predictive}, and more recently guided by systematic construction principles \cite{lebeda2025meta}. This framework naturally organizes semilocal density functionals according to their ingredients: LDA/LSDA depend only on the electron density \cite{kohn1965self,von1972local}, GGAs incorporate its gradient \cite{perdew1996generalized,becke1988density,perdew1992atoms,constantin2011semiclassical,perdew2008restoring,vela2012improved}, and meta-GGAs further include the kinetic energy density and/or Laplacian of electron density \cite{tao2003climbing,perdew2009workhorse,sun2012communication,zhao2006new,del2012new,sun2015strongly,bartok2019regularized,aschebrock2019ultranonlocality,furness2020accurate,lak_functional}. Beyond semilocal approximations, hybrid functionals such as HSE \cite{hse03,hse03E,krukau2006influence} can improve accuracy, albeit at significantly higher computational cost.

Despite this systematic hierarchy, the performance of semilocal density functionals for defect formation energies remains not fully understood. LSDA, although exact for the uniform electron gas, tends to overbind and predict too-short lattice constants \cite{sun2019density}, yet it often yields surprisingly accurate vacancy formation energies in metals. In contrast, GGAs such as PBE \cite{perdew1996generalized} improve atomization energies but predict too-long lattice constants and systematically underestimate defect formation energies, reflecting the well-known tradeoff between structural and energetic accuracy \cite{perdew2008restoring,sun2019density}. Meta-GGAs such as SCAN \cite{sun2015strongly} and \(r^2\)SCAN \cite{furness2020accurate} achieve broadly improved and transferable accuracy across bonding types, but their performance for defects is mixed and not yet fully rationalized \cite{jana2018assessing,kaplan2022laplacian,sarkar2026revisiting,he2023assessment,patra2025vacancy,abbas2025theoretical}. For semiconductors and insulators, the TASK functional \cite{aschebrock2019ultranonlocality} achieves accurate band gaps for the correct physical reason \cite{lebeda2023right}: by capturing the ultranonlocality associated with the derivative discontinuity \cite{perdew1982density}. While this also improves response properties \cite{aschebrock2023exact}, TASK's accuracy for chemical bonding is limited \cite{lebeda2022first}.

Motivated by these developments, the Lebeda--Aschebrock--K\"ummel (LAK) functional \cite{lak_functional} extends the non-empirical meta-GGA framework by combining full constraint satisfaction with two additional construction principles that guide the functional's behavior between the exact limits and appropriate norms \cite{lebeda2025meta}. The first construction principle advances TASK’s strategy to capture the ultranonlocality associated with the derivative discontinuity \cite{lebeda2023right,lak_functional}, while the second one enables the description of intermediate-range noncovalent interactions \cite{lebeda2025meta}. To facilitate the combination of both construction principles in a nonempirical meta-GGA, LAK additionally takes advantage of a more balanced treatment of the gradient expansion \cite{lak_functional}. As a result, LAK achieves HSE-level accuracy for semiconductor band gaps, SCAN-level accuracy for bonding, and describes noncovalently bound systems near equilibrium without a dispersion correction \cite{lebeda2025meta}. Moreover, LAK was shown to capture challenging phenomena such as \(f\)-electron localization in the iso-structural \(\alpha\)–\(\gamma\) transition of Ce \cite{giri2025isostructural}.

The study of defects is a central application of DFT, providing microscopic insight that guides materials design across applications such as solar cells, catalysis, and quantum information \cite{freysoldt2014first,evarestov2012quantum,ivady2018first,zhang2025transition}. A prototypical benchmark is the formation energy of monovacancies in fcc metals, where extensive studies have shown that GGAs underestimate formation energies while LSDA performs unexpectedly well \cite{korhonen1995vacancy,soderlind2000first,carling2000vacancies,mattsson2002calculating,delczeg2009assessing,nazarov2012vacancy}, and hybrid functionals such as HSE have been tested only in limited cases \cite{xing2014vacancy}. Experimental determinations themselves show significant scatter \cite{erhart1991atomic,nazarov2012vacancy,xing2014vacancy}, underscoring the need for reliable theoretical predictions.

In this work, we pursue two complementary objectives. First, we assess the performance of the recently developed LAK functional for defect formation energies in both metals and semiconductors, motivated by its improved description of electronic structure. Second,
we use defect formation energies as a stringent probe to analyze how the ingredients of semilocal density functionals control their performance in strongly inhomogeneous environments. By systematically comparing LDA, GGA, and meta-GGA functionals (SCAN, \(r^2\)SCAN, and LAK), we identify how different ingredients behave in the critical regions associated with defects, elucidating the origin of their successes and failures across materials. Specifically, we present a systematic assessment of monovacancies in fcc metals and analyze the behavior of functional ingredients in these systems. To complement this, we investigate a prototypical semiconductor challenge: diamond interstitial defects in Si. 
Together, these results provide new insight into the role of functional ingredients in semilocal DFT and offer guidance for future functional development.

\section{Computational Methods}
The calculations were carried out using the plane-wave–based Vienna $ab$ $initio$ Simulation Package (\textsc{vasp} 6.5.1) \cite{kresse1996efficient,kresse1993ab}, employing the projector augmented-wave (PAW) pseudopotential method \cite{kresse1999ultrasoft}. The density functionals LDA \cite{dirac1930note, ceperley1980ground,PhysRevB.23.5048}, PBE \cite{perdew1996generalized}, SCAN \cite{sun2015strongly}, $r^2$SCAN \cite{furness2020accurate}, and LAK \cite{lak_functional,lebeda2025meta} were applied to all studied systems. For Si-diamond we also used the hybrid HSE06 \cite{krukau2006influence}.
Next, we describe the methodology for each of the systems we studied.

\subsection{Fcc metals}\label{subsec:fcc_metal_methods}
For the fcc metals, we performed calculations with fully relaxed structures, important for applications, and at unrelaxed structures with experimental lattice constants \cite{hao2012lattice,sun2011self} to facilitate a comparison of the semilocal ingredients across different functionals.
For a defective system, using the unrelaxed structure means removing the defective atom from the pristine system with experimental lattice constants. However, this does not correspond to a measurable structure for the defective system since it neglects the structural changes induced by the defect. Thus, for practical considerations the results for relaxed structures should be considered.

To obtain the relaxed structures, we first fully relaxed the conventional bulk unit cell using a regular $k$-point mesh of 7$\times$7$\times$7 and a cut-off energy of 700 eV. Then, we converged the total energy to $10^{-6}$ eV and the atomic forces to 0.001 eV/\si{\angstrom}. In addition, we used the conjugate gradient algorithm along with the Methfessel-Paxton method of first order (ISMEAR = 1) and a smearing of 0.2 eV.

With the relaxed bulk conventional structure, we built two 2$\times$2$\times$2 supercells: the pristine and the defective. The latter includes a monovacancy, i.e. taking out one atom from the pristine supercell.
Then, we fully relaxed both systems using a regular $k$-point mesh of 17$\times$17$\times$17, and the same settings used for the conventional bulk cell. In addition, to obtain the total energy, we used the relaxed supercells and performed a single point calculation using the tetrahedron method with Bl\"ochl corrections. 
The formation energy $E_\mathrm{f}$ of a monovacancy is computed as follows:  
\begin{equation}
   E_\mathrm{f} =  E^{\mathrm{vacancy}} - \frac{n - 1}{n} E^{\mathrm{pristine}}~,
\end{equation}
where $E_\mathrm{vacancy}$ is the total energy of the defective supercell, $E_\mathrm{pristine}$ is the total energy of the pristine supercell and $n$ is the total number of atoms in the pristine supercell. This definition encompasses the surface, curvature, and higher-order contributions to the energy associated with the creation of the curved inner surface around the vacancy.

We note that for some supercells, the algorithm search ZBRENT, as part of the conjugate gradient algorithm (IBRION = 2), can fail. Hence, the RMM-DIIS (IBRION=1) method and an additional support grid for the evaluation of the augmentation charges (ADDGRID=.TRUE.) were used. 

For computing the defect formation energies of unrelaxed structures, we perform a one-ionic step calculation with all atomic and lattice degrees of freedom fixed. The $k$-point mesh is 17$\times$17$\times$17 and we converged the total energy to $10^{-6}$ eV and the atomic forces to 0.001 eV/\si{\angstrom}. Moreover, we used the conjugate gradient algorithm along with the Methfessel-Paxton method of first order (ISMEAR = 1) and a smearing of 0.2 eV. 

\subsection{Si-diamond interstitials}\label{sec:si_diamond}
To study Si-diamond interstitials, we need the total energies of the reference bulk structure, and the pristine and defective supercells. For bulk calculations, we used a regular $k$-point mesh of 15$\times$15$\times$15 and a cut-off energy of 700 eV. The total energy converged to $10^{-6}$ eV and the atomic forces to 0.005 eV/\si{\angstrom}. In addition, we used the conjugate gradient algorithm along with Gaussian smearing (ISMEAR = 0) of 0.05 eV width. The pristine supercell consists of 64 Si atoms, whereas the defective ones contain 65 atoms. The supercells were fully relaxed using the same settings as for the bulk, with exception of the $k$ points. Specifically, a regular $k$-point mesh of 4$\times$4$\times$4 for all supercells except for the tetrahedral (T) defect was used. For T, we used a 7$\times$7$\times$7 mesh based on convergence tests. Furthermore, with the relaxed structures, we performed a single point calculation using the tetrahedron method with Bl\"ochl corrections. The defect formation energy is obtained as follows:
\begin{equation}
    E_\mathrm{f} = E^\mathrm{defective} - E^\mathrm{pristine} - \mu_\mathrm{Si}~,
    \label{Eform_si}
\end{equation}
where $E^\mathrm{defective}$ is the total energy of the defective supercell, $E^\mathrm{pristine}$ is the total energy of the pristine supercell and $\mu_\mathrm{Si}$ is the chemical potential of silicon and corresponds to the total energy of bulk Si per atom. This definition encompasses the interaction between bulk silicon and a self-interstitial atom.

\subsection{Calculations of the density functionals' ingredients}
In order to investigate the meta-GGA ingredients for both fcc metals and Si-diamond,
we performed a one-ionic step calculation with all atomic and lattice degrees of freedom fixed and wrote the all-electron charge density to a file (LAECHG=.TRUE.) with $160\times 160 \times 160$ points for both the coarse (NGX, NGY, NGZ) and finer (NGXF,NGYF, NGZF) FFT grids. We used the same settings of cut-off energy, $k$ points, energy and force tolerances as described in Sections \ref{subsec:fcc_metal_methods} and \ref{sec:si_diamond}. In contrast to the fully relaxed calculations which employed the standard pseudopotentials recommended by \textsc{vasp}, we used the GW pseudopotentials for computing the ingredients using the unrelaxed structures. This applies for all metals except for the case of SCAN on Al, since the latter did not converge with the GW pseudopotential. Our rationale for using the GW pseudopotentials here is that they treat more electrons explicitly, allowing for reliable plots of the meta-GGA ingredients in a larger region. To validate this approach, we compare the obtained ingredients for Pt with those of the all-electron code \textsc{band} \cite{band1,band2,band3,band4,band2023.1} in the Supplemental Material \cite{supplemental}, see Fig.~S9.


\section{Results and Discussion}\label{sec:results}

\subsection{Formation energy of monovacancies in fcc metals}\label{sec:fcc_metals}
\begin{figure}[h!]
    \centering
    \includegraphics[width=1\linewidth]{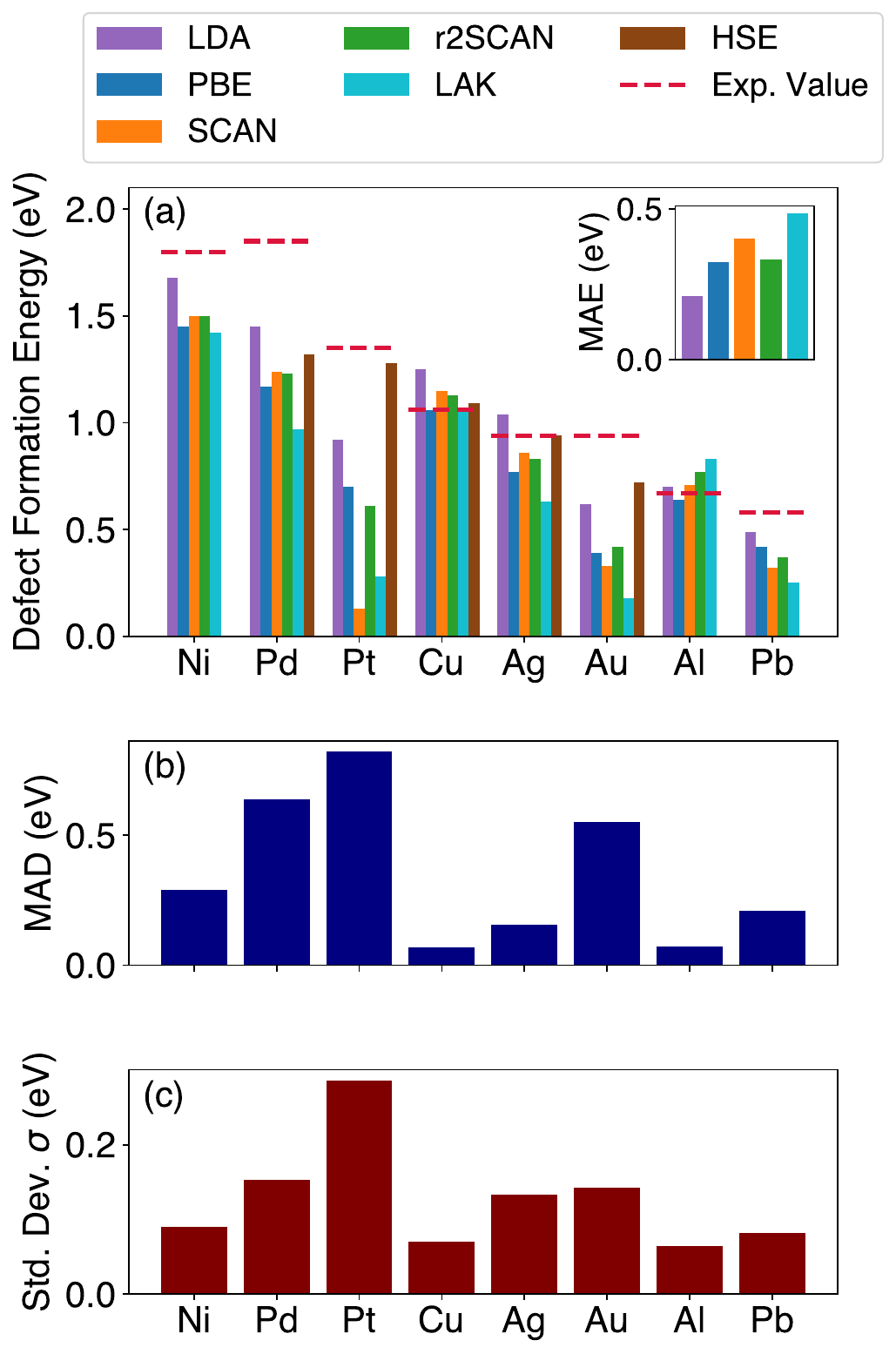}
 \caption{(a) Monovacancy defect formation energies for fcc metals, with the mean absolute errors of the DFAs shown in the inset. The dashed lines denote the experimental values. Experimental numerical values along with their references and related remarks are provided in the Supplemental Material \cite{supplemental}. For Ni and Pd we considered ferromagnetic (FM) spin arrangements. (b) Mean absolute deviation (MAD) across the considered semilocal DFAs for each metal. (c) Standard deviations for each metal, again excluding HSE. Fully relaxed structures are used. HSE values are taken from Ref.~\onlinecite{xing2014vacancy}.}
 \label{fcc_figure_theoretical}
\end{figure}

Figure \ref{fcc_figure_theoretical} shows our results for fully relaxed structures. For the detailed data and additional comments, refer to the Supplemental Material \cite{supplemental}, see Table S2.


We observe the following trends:

\begin{enumerate}
    \item[i)] In Fig.~\ref{fcc_figure_theoretical}(a), all predicted defect formation energies decrease as the atomic radius increases for elements within the same column of the periodic table. The best experimental values, however, do not strictly follow this trend.

    As noted in Refs.~\onlinecite{xing2014vacancy, glensk2014breakdown}, the commonly accepted experimental values of defect formation energies may require revision due to the pronounced temperature dependence of the Gibbs formation energy. In particular, the revised values for Cu and Ag reported in Refs.~\onlinecite{glensk2014breakdown, xing2014vacancy} recover the same trend predicted by the DFAs, with Ag exhibiting a lower defect formation energy than Cu. For other systems, additional uncertainties remain. For instance, in the case of Au, no (non-Arrhenius) local Gr\"uneisen theory (LGT) correction was applied due to the lack of reliable experimental data, although such a correction is expected to reduce the reported formation energy and bring it closer to the DFA trend. More generally, the availability and reliability of experimental data for vacancy formation energies remain limited (see Supplemental Material \cite{supplemental}).

    These observations have important implications for previous studies that rely on experimental reference values, including approaches based on surface corrections and the AM05 density functional~\cite{armiento2005functional,mattsson2008am05, mattsson2009implementing}. For comparison, we also include HSE results from Ref.~\onlinecite{xing2014vacancy} for Pd, Pt, Cu, Ag, and Au, as shown in Fig. \ref{fcc_figure_theoretical}(a) for fully relaxed structures. For Cu and Ag, HSE yields defect formation energies in close agreement with the revised experimental values. For Pt, a slight underestimation is observed, while for Au and Pd the underestimation is more pronounced. We also note that the HSE results were obtained by optimizing the fraction of exact exchange for each system. While this approach improves agreement with experiment, it introduces a degree of empiricism and significantly increases the computational cost, in contrast to semilocal DFAs, which are more computationally efficient and broadly applicable.
    \item[ii)] Following the previous observation, for elements within the same column of the periodic table (e.g., Ni, Pd, Pt and Cu, Ag, Au), DFAs' predictions for elements with small atomic radius are more accurate and have relatively small spread among the functionals. In contrast, for elements with large atomic radius, all DFAs underestimate the monovacancy formation energies and the spread increases. As a measure of the difficulty for describing each system, we report for each system the mean absolute deviation (MAD) among the considered DFAs with respect to the experimental values in Fig.~\ref{fcc_figure_theoretical}(b). Similarly, as a measure of the spread among the DFAs, we show for each system the standard deviation of the considered DFA's results in Fig.~\ref{fcc_figure_theoretical}(c). The DFAs show excellent performance for the Cu and Al vacancy formation energies, both having the lowest MAD and standard deviation. In contrast, Au, Pd, and Pt are the most difficult systems with the largest MAD and standard deviation. 
    \item[iii)] As a measure of the overall accuracy of each DFA we report the mean absolute error (MAE) with respect to the experimental values as insets in Fig.~\ref{fcc_figure_theoretical}(a). Of all DFAs considered, LDA provides the closest agreement with experiment overall, whereas LAK is the least accurate. More precisely, the MAE increases as we go from LDA to PBE, $r^2$SCAN, SCAN, and LAK. For most of the metals, the defect formation energies obtained with LDA are the highest, while the ones given by LAK are the lowest. Remarkably, we observe the opposite trend for Al, where LAK and $r^2$SCAN give the highest defect formation energies. We rationalize this opposite trend for Al from the underlying semilocal ingredients in Section \ref{sec:metals analysis}.

    \item[iv)] For Pt, Au, and Pb, we also calculated the defect formation energies considering Spin Orbit Coupling (SOC) using PBE and the unrelaxed structures. We observe no qualitative change in the defect formation energies with respect to the results without SOC. For the numerical values refer to the Supplemental Material \cite{supplemental}.
    
\end{enumerate}

\subsection{Formation energy of interstitial defects in diamond Si} \label{sec:Si_interstitials}

\begin{figure}[h]
    \centering
\includegraphics[width=1\linewidth]{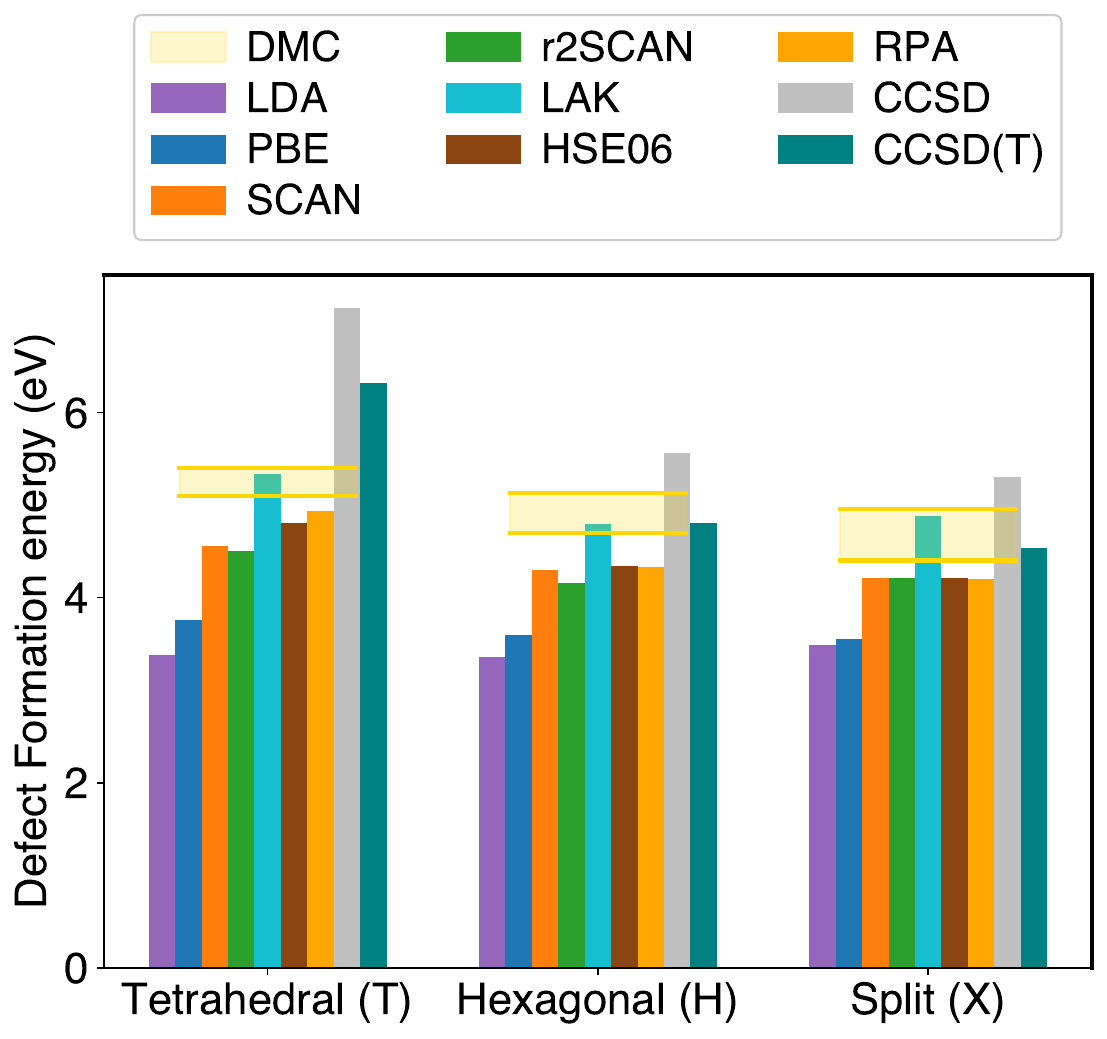}
 \caption{Formation energies of the interstitial defects in Si-diamond for various functionals calculated with the relaxed structures compared to DMC. The DMC range values are taken from Refs.~\onlinecite{parker2011accuracy,leung1999calculations,batista2006comparison}.} 
 \label{si-interstitials}
\end{figure}

In this section we present our results for the defect formation energy of the split (X), Hexagonal (H), and Tetrahedral (T) self interstitials of Si-diamond.
Figure \ref{si-interstitials} shows the defect formation energies for all DFAs along a range of published diffusion Monte Carlo (DMC) values highlighted in yellow as a benchmark. For the latter, we used the lowest and highest values from several studies employing the DMC method \cite{leung1999calculations,batista2006comparison,parker2011accuracy}. Experimental estimates of the self-interstitial formation energy in Si are typically inferred from diffusion measurements and lie in the range of approximately 4$-$5 eV \cite{fahey1989point,ural1999self, bracht1998silicon,leung1999calculations}, although these values depend on the separation of formation and migration contributions. More direct theoretical benchmarks, such as DMC, provide formation energies within a similar range and we consider this approach as our benchmark.
For the detailed data, refer to Table S1 in the Supplemental Material \cite{supplemental}.

Furthermore, we show in Fig.~\ref{si-interstitials} the results from other methods such as RPA \cite{kaltak2014cubic}, CCSD \cite{salihbegovic2023formation}, and CCSD(T) \cite{salihbegovic2023formation} to make a comparison with our DFT results. Among all studied methods, LAK is the only one that consistently achieves defect formation energies within the DMC range. This is remarkable because LAK provides this DMC-level accuracy at very affordable semilocal computational cost.

The accuracy of the other density functionals follows in the order HSE, SCAN and $r^2$SCAN, PBE, and LDA. Whereas HSE, SCAN, and $r^2$SCAN show a consistent but slight underestimation with respect to the lowest DMC value, PBE and LDA severely underestimate the defect formation energies.

For the RPA, CCSD, and CCSD(T) methods, we note that RPA yields defect formation energies very close to the HSE, CCSD systematically overestimates them, and CCSD(T) yields values inside the DMC range for the X- and H-interstitial defects. However, for the T-interstitial defect CCSD(T) shows a significant overestimation. Given the high computational cost of RPA and CCSD methods, their mixed accuracy underscores that LAK's performance is exceptional.

A useful comparison is provided by examining the predicted energetic ordering of the interstitial configurations across methods. Most semilocal and hybrid functionals (PBE, SCAN, and HSE06), as well as coupled-cluster approaches (CCSD and CCSD(T)), and RPA, predict the split (X) interstitial as the lowest-energy configuration, typically followed by H and T. In contrast, $r^2$SCAN and LAK favor the H configuration, followed by X and T.

The available DMC results do not provide a uniform picture: Reference \onlinecite{batista2006comparison} reports formation energies in the order X $<$ T $<$ H, Ref.~\onlinecite{leung1999calculations} finds H $<$ X $<$ T, and Ref.~\onlinecite{parker2011accuracy} obtains X $<$ H $<$ T. Although two of these studies agree on X as the most stable configuration, the overall lack of consensus indicates that the formation energies of the competing interstitials are very close. These mixed results suggest that the relative stability of interstitial configurations lies within the typical accuracy limits of current methods, making it challenging to assign a definitive ground state and underscoring the need for highly accurate reference data.

\subsection{Physical interpretation of defect formation trends of metals}

Our results are in line with general trends for DFAs: There is a clear improvement in accuracy from LSDA to GGA to meta-GGA for atoms, molecules, semiconductors, and insulators. However, for some properties of metals, LSDA and GGA are more accurate. We attribute this to the perfect long-range screening in metals that makes the exchange-correlation hole short-ranged, as it is in LSDA and PBE but not necessarily in the meta-GGAs SCAN, r$^2$SCAN, and LAK.

Within the macroscopic liquid drop model \cite{perdew1991liquid}, the formation energy of a spherical void of radius $R$ in a bulk solid is  
\begin{equation}
E_\mathrm{f} = \sigma4\pi R^2-\gamma2\pi R+\delta~,
\end{equation}

where $\sigma$ is the surface energy, $\gamma$ is the curvature energy, and $\delta$ is a constant. Within the stabilized jellium model \cite{perdew1990stabilized} for a metal of any average valence electron density, $\sigma$ and $\gamma$ are positive, $\delta$ is small, and the liquid drop model is usefully accurate down to the microscopic monovacancy radius \cite{ziesche1994spherical}, where the surface term is still the largest term.  If the ordinary jellium model (stable only near the average valence electron density of Na) is a guide, the surface energy is accurate in LSDA (due to a strong error cancellation between exchange and correlation), too low in PBE GGA, and accurate again in SCAN meta-GGA (with little error cancellation)~\cite{patra2017properties}. Thus, the reduction in the monovacancy formation energy in a real metal from LSDA to PBE is probably attributable to the reduction in surface energy, while the reduction in vacancy formation energy from LSDA to SCAN/r$^2$SCAN is probably due to an overestimation of the curvature energy of a metal by the latter density functionals.

\section{Rationalizing the trends in defect formation energies from the semilocal ingredients}


In this section, we rationalize the trends we described in the previous section from an analysis of the semilocal ingredients $r_s$, $s$, and $\alpha$. Since the ingredients analysis is easier for the fcc metals, we discuss them first. To explain the trends we described in Section \ref{sec:fcc_metals}, we analyze the behavior of the semilocal ingredients along a path from an atom neighboring the vacancy to an atom or vacancy, corresponding to the pristine or defective systems, respectively. For the Si interstitials, the analysis is more involved, since the defects are not simply vacancies but additional atoms that change the crystal structure. Thus, to rationalize our observations of Section \ref{sec:Si_interstitials}, we analyze the behavior of the semilocal ingredients on a 2D section through the crystal that contains the defect.

\subsection{Technical background}



The xc energy can be written as
\begin{equation}
    E_\mathrm{xc} = \int d\mathbf{r} \; e_\mathrm{xc}(\mathbf{r})~,
\end{equation}

where $e_\mathrm{xc}(\mathbf{r})$ is the negative xc energy density.
Semilocal approximations to $e_\mathrm{xc}(\textbf r)$ are classified as LDA, GGA, or meta-GGAs depending on the semilocal ingredients used.
Specifically, in terms of dimensionless ingredients, the LDA includes the Wigner-Seitz radius $r_s(\mathbf{r}) = \left(4\pi n(\mathbf{r})/3\right)^{-1/3}$, the GGA adds the reduced density gradient ${s(\mathbf{r}) = \vert\nabla n(\mathbf{r})\vert / [2(3\pi^2)^{1/3} n(\mathbf{r})^{4/3} ]}$, and meta-GGAs typically add the iso-orbital indicator $\alpha(\mathbf{r}) = \left(\tau(\mathbf{r}) - \tau_W(\mathbf{r})\right) / \tau_\mathrm{unif}(\mathbf{r})$, where $\tau(\mathbf{r}) = \sum_{\sigma,i}^\mathrm{occ} \left| \nabla \varphi_{i\sigma}(\mathbf{r}) \right|^2 /2 $ is the noninteracting kinetic energy density, $ {\tau_W(\mathbf{r}) = \vert\nabla n(\mathbf{r})\vert^2/8n(\mathbf{r})}$ is the iso-orbital limit of $\tau$, and ${\tau_\mathrm{unif}(\mathbf{r}) = (3/10)(3\pi^2)^{2/3} n(\mathbf{r})^{5/3}}$ is the homogeneous electron gas limit of $\tau$. The ingredients $r_s$, $s$, and $\alpha$ are called semilocal because they are constructed from $n(\textbf r)$, $\nabla n(\textbf r)$, and $\nabla \varphi_{i\sigma}(\textbf r)$, which are all local in $\textbf r$ and available in a Kohn-Sham or generalized Kohn-Sham calculation. $\alpha$ has been shown to identify different chemical bonds with $\alpha =0$ for single covalent bonds, $\alpha \sim 1$ for metallic bonds, and $\alpha \gg 1$ for non-covalent bonds \cite{sun2013density}. 

To make the differences and trends in the ingredients more transparent, we consider $\Delta r_s=r_{s}^\mathrm{def} - r_{s}^\mathrm{pris} $, $\Delta s=s^\mathrm{def}-s^\mathrm{pris}$, and $\Delta \alpha = \alpha^\mathrm{def} - \alpha^\mathrm{pris}$, i.e., the difference between the ingredient in the defective structure and in the pristine one.


To analyze the change of $e_\mathrm{xc}(\textbf r)$ between the defective and pristine systems, we define the ratio
\begin{equation}
   R_\mathrm{xc}(\textbf{r}) = \frac{e_\mathrm{xc}^\mathrm{def}(\textbf{r})}{e_\mathrm{xc}^\mathrm{pris}(\textbf{r})}~.
\end{equation}

$R_\mathrm{xc} > 1$ corresponds to  $|e_\mathrm{xc}^\mathrm{def}|>|e_\mathrm{xc}^\mathrm{pris}|$ and signals that the defective system is energetically favored, whereas $R_\mathrm{xc} < 1$ implies $|e_\mathrm{xc}^\mathrm{def}|<|e_\mathrm{xc}^\mathrm{pris}|$, i.e., the defective system is disfavored. To compare the behavior of $R_\mathrm{xc}$ among the DFAs, we calibrate $R_\mathrm{xc}$ with respect to that of LDA,
\begin{equation}
    \Delta R_\mathrm{xc}^\mathrm{DFA} = R_\mathrm{xc}^\mathrm{LDA} - R_\mathrm{xc}^\mathrm{DFA}~.
\end{equation}

Note that we base $R_\mathrm{xc}$ on $e_\mathrm{xc}(\textbf r)$ and not the often studied enhancement factor $F_\mathrm{xc}(\textbf r) = e_\mathrm{xc}(\textbf r) / e_\mathrm{xc}^\mathrm{LDA}(\textbf r)$ because differences in $F_\mathrm{xc}(\textbf r)$ become large close to the vacancy, although this region is not energetically important because the density is extremely small there. Moreover, we define $\Delta R_{xc}$ as $R_\mathrm{xc}^\mathrm{LDA} - R_\mathrm{xc}^\mathrm{DFA}$ (rather than $R_\mathrm{xc}^\mathrm{DFA}-R_\mathrm{xc}^\mathrm{LDA}$) to make the interpretation of the insets in Figs. \ref{fig:ingredients_Al_Ni_Pd_Pt} and \ref{fig:ingredients_Cu_Ag_Au_Pb} more straightforward: a negative $\Delta R_{xc}$ indicates smaller defect formation energies.

\subsection{Analysis of the semilocal ingredients for fcc metals}\label{sec:metals analysis}


\begin{figure}[htb]
    \centering
    \includegraphics[width=1\linewidth]
    {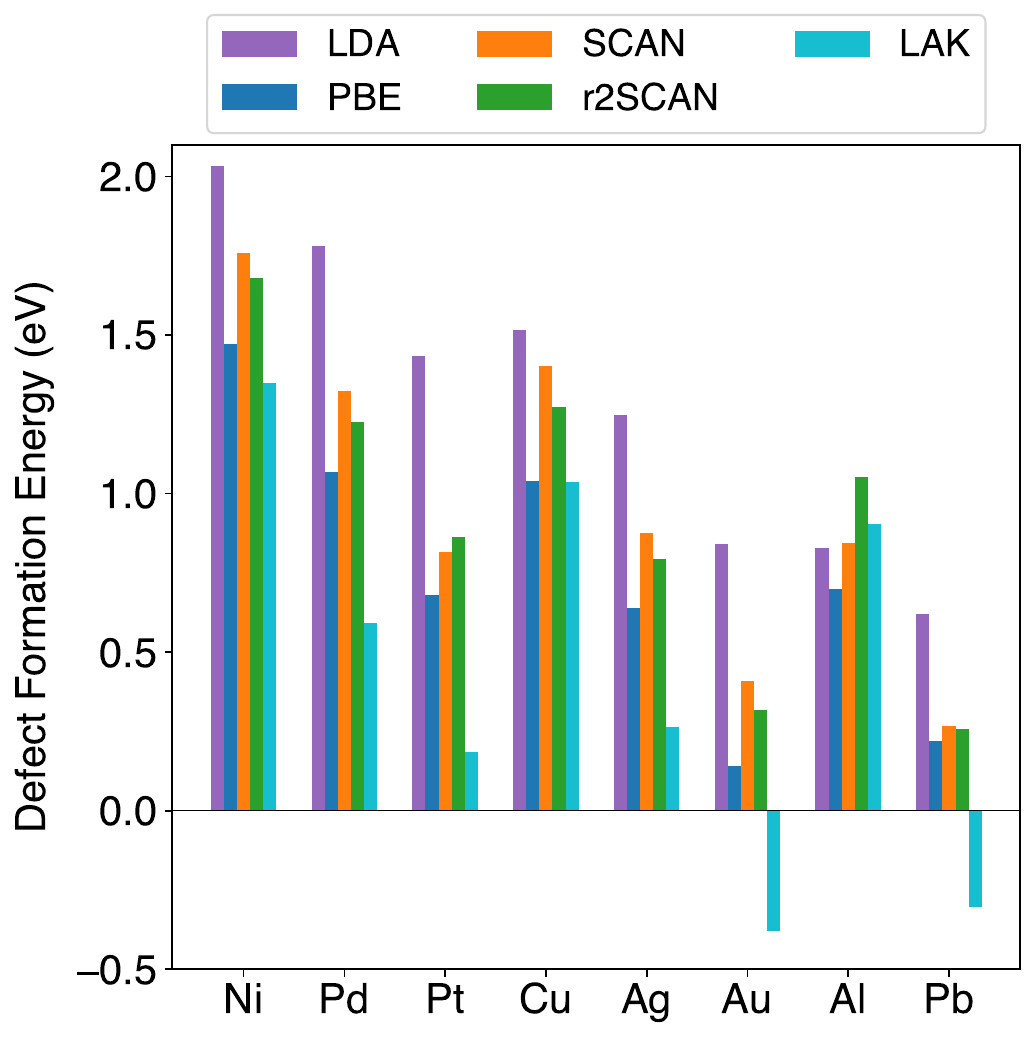}
 \caption{Monovacancy defect formation energies of fcc metals. Unrelaxed structures using experimental lattice constants and fixed atomic positions are used. This figure is intended for the ingredients analysis; for other purposes refer to Fig.~\ref{fcc_figure_theoretical}.}
 \label{fcc_figure_experimental}
\end{figure}

\begin{figure*}[htb]
  \includegraphics[width=0.9\textwidth]
  {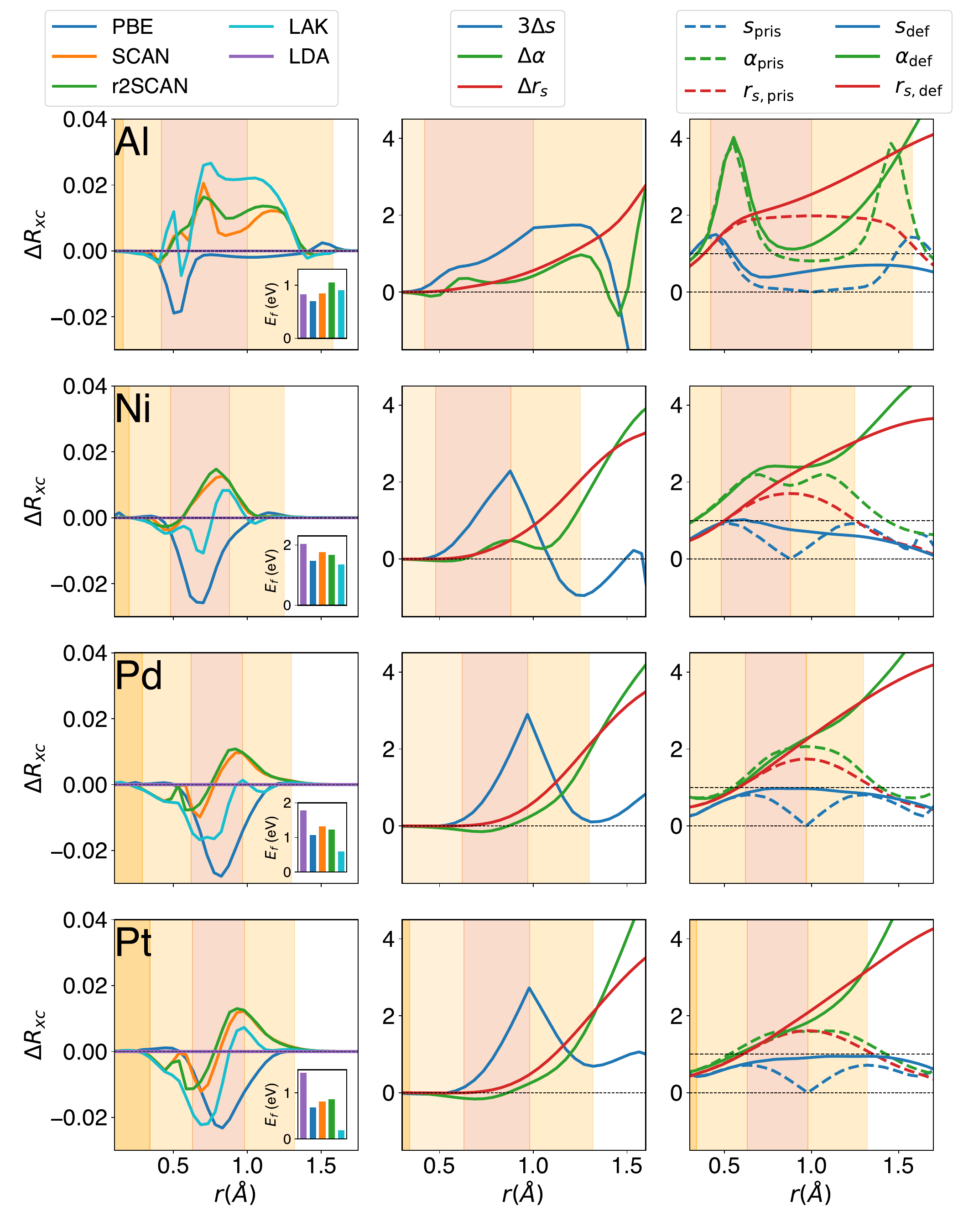}
  \caption{Ingredients analysis for Al, Ni, Pd, and Pt. Right: the semilocal ingredients $r_s$, $s$, and $\alpha$. Middle: $\Delta r_s$, $3\Delta s$, and $\Delta \alpha$,  where $\Delta x := x_\mathrm{defective} - x_\mathrm{pristine}$. Left: $\Delta R_\mathrm{xc} = R_\mathrm{xc}^\mathrm{LDA} - R_\mathrm{xc}^\mathrm{DFA}$, where $R_\mathrm{xc} = e_\mathrm{xc}^\mathrm{def}/e_\mathrm{xc}^\mathrm{pris}$. In addition, each plot shows an inset with the defect formation energies obtained with the experimental lattice constants.
    For all the plots, the path is chosen as the shortest path from an atom to the vacancy; one nucleus is on the left and the vacancy (or a second nucleus) is on the right. Several regions are highlighted: core (orange), dominant (pink), important (light orange), and vacancy (white). For definitions of these regions, see main text. Experimental lattice constants are used. Note that $\Delta R_\mathrm{xc}$ depends not only on the difference in ingredients but also critically on the value of the density. This is why the peak in $\Delta R_\mathrm{xc}$ for PBE is always shifted towards the nucleus compared with the peak in $\Delta s$.}
    \label{fig:ingredients_Al_Ni_Pd_Pt}
\end{figure*}

\begin{figure*}[htb]
  \includegraphics[width=0.9\textwidth]
  {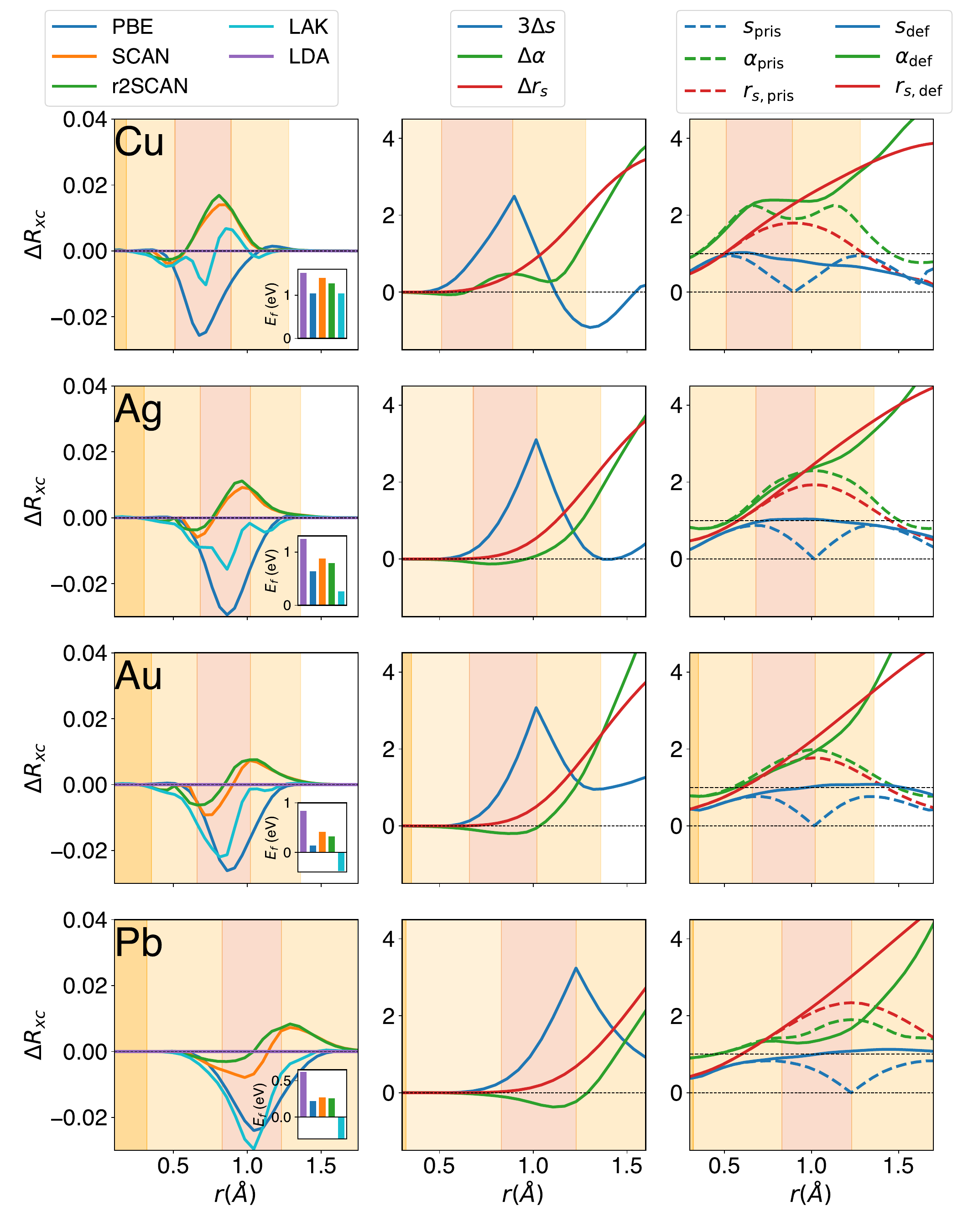}
  \caption{Ingredients analysis for Cu, Ag, Au, and Pb. Right: the semilocal ingredients $r_s$, $s$, and $\alpha$. Middle: $\Delta r_s$, $3\Delta s$, and $\Delta \alpha$,  where $\Delta x := x_\mathrm{defective} - x_\mathrm{pristine}$. Left: $\Delta R_\mathrm{xc} = R_\mathrm{xc}^\mathrm{LDA} - R_\mathrm{xc}^\mathrm{DFA}$, where $R_\mathrm{xc} = e_\mathrm{xc}^\mathrm{def}/e_\mathrm{xc}^\mathrm{pris}$. In addition, each plot shows an inset with the defect formation energies obtained with the experimental lattice constants.
    For all the plots, the path is chosen as the shortest path from an atom to the vacancy; one nucleus is on the left and the vacancy (or a second nucleus) is on the right. Several regions are highlighted: core (orange), dominant (pink), important (light orange), and vacancy (white). For definitions of these regions, see main text. Experimental lattice constants are used. Note that $\Delta R_\mathrm{xc}$ depends not only on the difference in ingredients but also critically on the value of the density. This is why the peak in $\Delta R_\mathrm{xc}$ for PBE is always shifted towards the nucleus compared with the peak in $\Delta s$.}
    \label{fig:ingredients_Cu_Ag_Au_Pb}
\end{figure*}

Figure \ref{fcc_figure_experimental} shows the defect formation energies of semilocal DFAs for the unrelaxed structures for which we calculated the semilocal ingredients. Although the qualitative trends are similar as for the relaxed structures, for the ingredients analysis the quantitative differences are important in order to understand how changes in the underlying ingredients between different systems lead to the respective defect formation energies.

Figures \ref{fig:ingredients_Al_Ni_Pd_Pt} and \ref{fig:ingredients_Cu_Ag_Au_Pb} show the results for Al, Ni, Pd, and Pt, and Cu, Ag, Au, and Pb, respectively, with both sets of metals going from smaller to higher atomic number. Figures \ref{fig:ingredients_Al_Ni_Pd_Pt} and \ref{fig:ingredients_Cu_Ag_Au_Pb} are divided into three panels: For each metal, the right hand side shows the semilocal ingredients $r_s$, $s$, and $\alpha$; the middle panel shows $\Delta r_s$, $\Delta s$, and $\Delta \alpha$ (difference in ingredients between pristine and defective system); and the left hand side shows $\Delta R_\mathrm{xc}$ (difference between $R_\mathrm{xc}^\mathrm{LDA}$ and $R_\mathrm{xc}^\mathrm{DFA}$). Each frame covers the shortest path from the vacancy to a neighboring atom in real space, i.e., the path along the chemical bond of the pristine system. In each frame, the left hand side starts around the outer core of one atom and the right hand side ends near the vacancy (or a closest atom in the pristine system).

From Figs.~\ref{fig:ingredients_Al_Ni_Pd_Pt} and \ref{fig:ingredients_Cu_Ag_Au_Pb}, we observe several trends: 
\begin{enumerate}
    \item[i)] As expected, $r_s$ of the defective system (solid red line in the right panels) increases monotonically when going from the atom (left) to the vacancy (right), reflecting the monotonic decrease of the electron density. The same holds for $\Delta r_s$ (red line in the middle panels) .
    
    \item[ii)] The reduced density gradient $s$ of the pristine system (dashed blue line) vanishes at the bond center. In the simple metal Al, there is a large region of constant $r_s$ and small $s$ near the bond center, reflecting a region of homogeneous electron density. In contrast, $s$ in the transition metals drops quickly near the bond center, giving rise to a distinctive peak in $\Delta s$.
    
    \item[iii)] The iso-orbital indicator $\alpha$ of the pristine system (dashed green line) can identify metallic bonds in simple metals and transition metals. In the simple metal Al, $\alpha$ approaches one near the bond center, whereas in the transition metals and the post-transition metal Pb, $\alpha$ takes values around two near the bond center. 
    
    \item[iv)] As a guidance for further analysis, we identify dominant and important regions using the $s$ distribution of the pristine systems, which contribute the most to differences in the defect formation energies between DFAs.
    The dominant region (pink) spans from the bond center (where $s=0$) to the next maximum in $s$ on the left hand side. In addition, there are two important regions (light orange). One spans from the bond center to the maximum of $s$ on the right hand side, and the other spans from the maximum of $s$ on the left hand side to the next minimum of $s$ on the left hand side. Further, there are unimportant regions. One is the core region (the darkest orange), which spans from the latter minimum in $s$ to the center of the nucleus. The other is the vacancy region (white), where the density in the defective system becomes very small. An important caveat in interpreting Figs.~\ref{fig:ingredients_Al_Ni_Pd_Pt} and \ref{fig:ingredients_Cu_Ag_Au_Pb} is that they overemphasize the bond center. When integrating the energy density, the bond center has a small weight, whereas the region of the first maximum in $s$ left to the bond center has a much higher weight. In Figs.~S10$-$S21 of the Supplemental Material \cite{supplemental}, we provide a complementary description of these observations that makes the trends of the ingredients more apparent. There, we show two-dimensional plots of the DFA ingredients and of the ratio $R_\mathrm{xc}$.

    \item[v)] $\Delta s$ exhibits a similar profile for all metals except Al, increasing monotonically and remaining positive up to the bond center. This trend indicates that $s_{\mathrm{defective}} > s_{\mathrm{pristine}}$ in the region that dominates the defect formation energy. The behavior of $s$ is straightforward to rationalize: by definition, $s$ vanishes at the bond center. Removing an atom causes the electron density to decay more rapidly in the vicinity of the vacancy, leading to a larger $s$ in the defective system. As a result, PBE stabilizes the defective structure relative to LDA, yielding smaller defect formation energies. 

    \item[vi)] Typically, $\Delta \alpha$ starts off negative near the nucleus, indicating increased orbital overlap in the pristine system.
    Then, $\Delta \alpha$ has a minimum and eventually becomes positive, reflecting increased orbital overlap and low density near the vacancy. The magnitude of the minimum in $\Delta \alpha$ increases with increasing atomic mass, most notably from 3d-transition metals (Ni, Cu) to 4d- and 5d-transition metals. We relate this to increasing orbital overlap between the 4d- and 5d-orbitals with 5s- and 6s-orbitals of the neighboring atom and associated s-d-hybridization when a transition metal bond forms. This is supported by the positions of the peaks of the d-orbital densities, which are at $r\approx0.5\si{\angstrom}$ in Cu, $r\approx0.9\si{\angstrom}$ in Ag, and $r\approx1.2\si{\angstrom}$ in Au \cite{blades2017evolution,desclaux1973relativistic}. This implies that the 3d-orbitals are very localized and close to the nucleus, which leads to negligible s-d orbital overlap and no significant s-d hybridization. In contrast, the 4d- and 5d-orbitals are more diffuse and contribute significantly s-d hybridization with the 5s and 6s orbitals, respectively, of the neighboring atom. In Au, additional 5d-5d orbital overlap between adjacent atoms further increases $\alpha$ in the pristine system and thus reducing $\Delta \alpha$ for $r\lesssim1$\AA.

    \item[vii)] $\Delta R_\mathrm{xc}$ has a characteristic shape for each DFA. For Al, $\Delta R_\mathrm{xc}$ of all DFAs differ from the trends persistent in the other metals. Regarding PBE, it presents a well-like shape that becomes wide as we go from smaller to larger atomic radius metals. On the other hand, $\Delta R_\mathrm{xc}$ of SCAN and r$^2$SCAN have very similar shapes. SCAN's $\Delta R_\mathrm{xc}$ is always slightly more positive than r$^2$SCAN. However, for all metals except Al and Pt, the defect formation energies of SCAN are slightly larger than r$^2$SCAN. Conversely, the shape of LAK's $\Delta R_\mathrm{xc}$ is reminiscent of PBE, albeit with clear deviations. As we go from smaller to larger atomic radius metals LAK's $\Delta R_\mathrm{xc}$ becomes larger and wider. Finally, for Al we note that the dominant region is wider than for the other metals. $\Delta R_\mathrm{xc}$ for PBE is always negative, whereas for the meta-GGAs it is predominantly positive. Also, if we compare with the defect formation energies, we see PBE's energy is smaller than the meta-GGAs. 


\end{enumerate}

Following the previous observations, we note that in the meta-GGAs, the dependence on $\alpha$ becomes important as $\alpha$ starts to differ at higher densities. Thus, we focus on $\alpha$ in the following and the trends observed in Figs.~S1 and S2. In the important region of Al between the core and the dominant regions, $\alpha$ becomes only slightly smaller in the defective system, while $\alpha$ becomes significantly larger in the dominant and the right hand side important region. 

Moving from Al to Ni, Pd, and Pt, the region of $\Delta\alpha < 0$ grows and the region of $\Delta\alpha > 0$ shifts toward the vacancy. As a consequence, in the left important region the meta-GGAs SCAN, $r^2$SCAN, and LAK, whose enhancement factors are all decreasing with $\alpha$ (at least in this parameter space), favor the defective system more than the pristine one in the transition metals. For SCAN and $r^2$SCAN, this difference is small and compensated by favoring the pristine system more in the right important region where $\Delta\alpha > 0$. However, for LAK this difference is significantly larger and less compensated in the right important region, due to the stronger dependence on $\alpha$ of its enhancement factor~\cite{lak_functional}. Consequently, the defect formation energy of LAK is significantly smaller. Since this effect increases from Ni over Pd to Pt, the difference in defect formation energies between LAK and the other functionals also increases from Ni over Pd to Pt, eventually leading to a very low defect formation energy of LAK for Pt. Similar explanations can be applied to Cu, Ag, and Au as well.


The different behavior of LAK is due to its different balancing of the contributions to the gradient expansion \cite{lak_functional}, as we detail for the example of Pt. In Pt, $\alpha$ between the defective and the pristine system starts to differ for $\alpha\gtrsim0.6$ and up to $\alpha\lesssim1.6$. Thus, the region of $\alpha\approx1$ is of particular importance for the defect formation energy of Pt. At $\alpha=1$, SCAN's dependence on $\alpha$ vanishes by construction to satisfy the gradient expansion with only contributions from $s$, i.e., from $\nabla n$. In contrast, LAK explicitly uses both $\nabla n$ and $\tau$, i.e., both $s$ and $\alpha$ to satisfy the gradient expansion. Moreover, LAK is based on the construction principle $\partial e_\mathrm{xc} / \partial \alpha > 0$, i.e., $e_\mathrm{xc}$ becomes less negative for larger $\alpha$. Since in the left important region of Pt $\alpha$ is larger in the pristine system than in the defective one, $e_\mathrm{xc}^\mathrm{pris}$  with LAK is less negative relative to $e_\mathrm{xc}^\mathrm{def}$  and 
$R_\mathrm{xc}$ is larger for LAK than for the other functionals. This corresponds to a more negative energy of the defective system and thus to a smaller defect formation energy of Pt with LAK, as we observe in Fig.~\ref{fcc_figure_experimental}.

In summary, we have shown why LAK predicts among the largest defect formation energy of Al but the smallest one for Pt: LAK's particularly strong dependence on $\alpha$ energetically favors systems with small values of $\alpha$.

\subsection{Analysis of the semilocal ingredients for interstitials in Si-diamond}\label{sec:si-diamond}

\begin{figure}[htb]
  \includegraphics[width=0.5\textwidth]
  {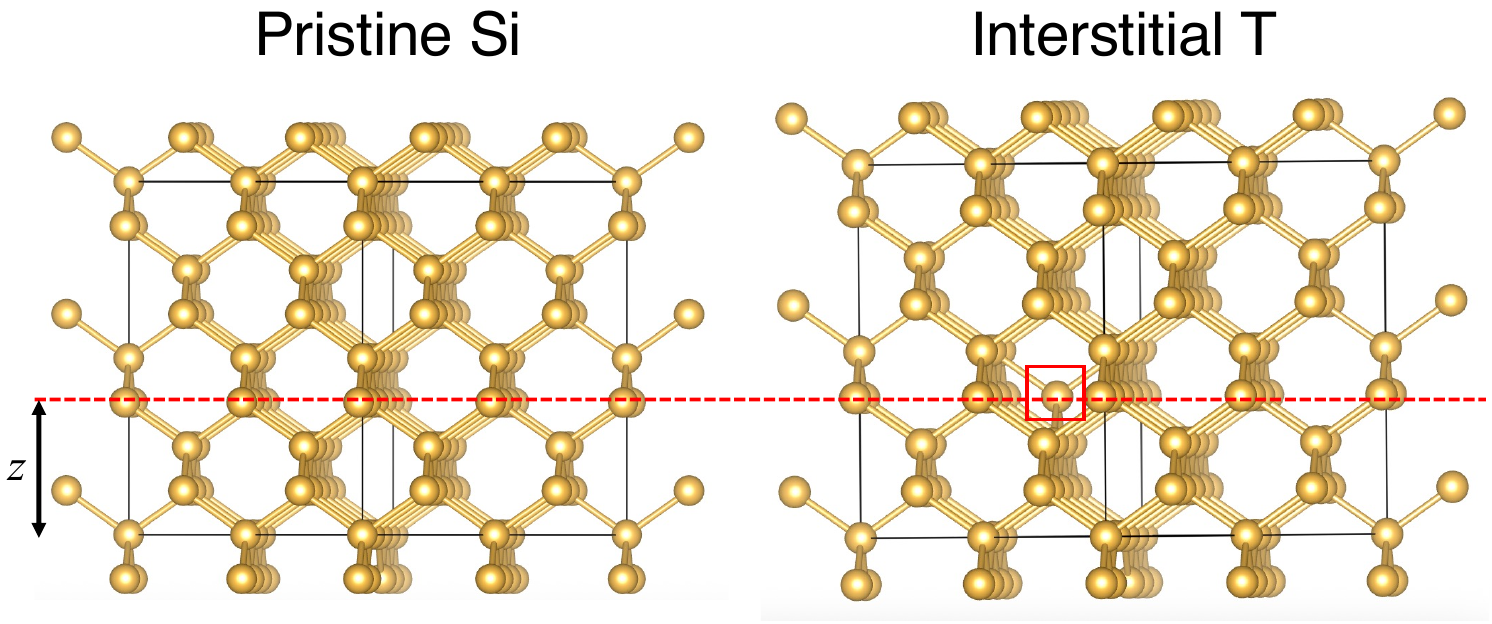}
  \caption{Side view of the pristine (left) and defective (right) Si-diamond structures. The box in red encloses the region with the defect, the red dashed line is the plane at which the cut for the ingredient analysis in 2D is made. $z$ is the height from the bottom of the supercell up to the plane that contains the interstitial.}
    \label{fig:2d_diamond_sketch_side}
\end{figure}

\begin{figure}[htb]
  \includegraphics[width=0.5\textwidth]
  {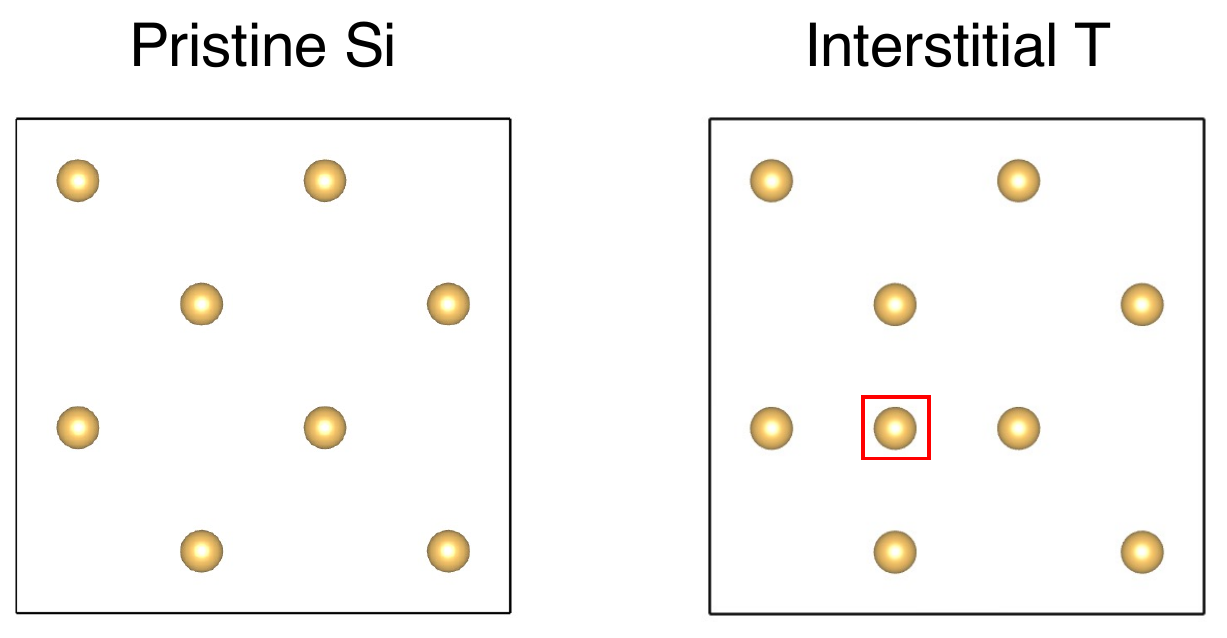}
  \caption{Plane view of the pristine (left) and defective (right) Si-diamond structures. The interstitial is enclosed in a red box. The 2D ingredients plots depict this plane. Atoms of other planes are not shown.}
    \label{fig:2d_diamond_sketch_plane}
\end{figure}

\begin{figure*}[htb]
  \includegraphics[width=\linewidth]
  {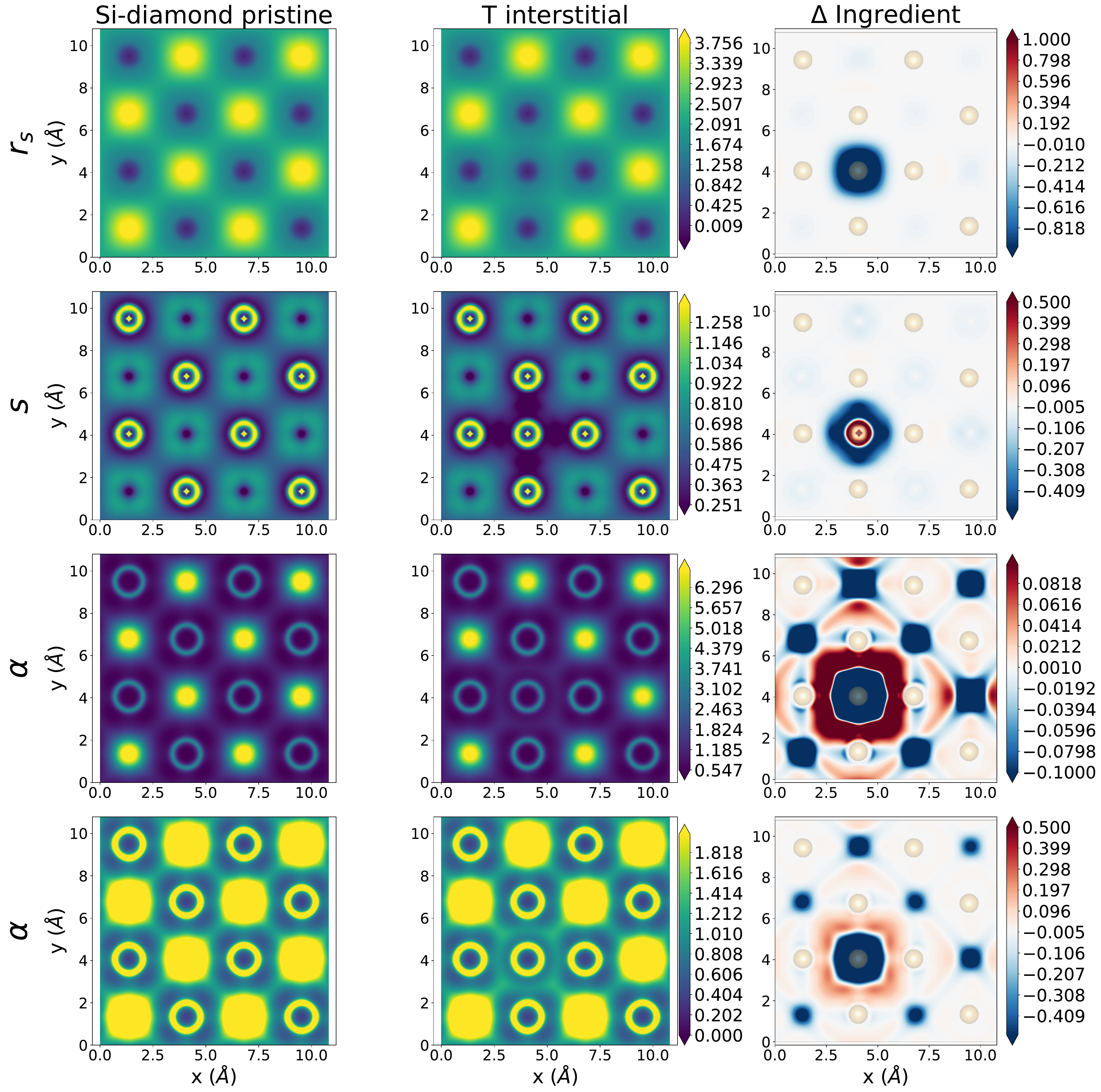}
  \caption{Ingredients in a real-space plane for the pristine Si-diamond (left column) and the T interstitial (middle column) for $r_s$  (1st row), $s$ (2nd row), and $\alpha$ (3rd and 4th rows). The 3rd column depicts the difference of the defective system ingredient with respect to the pristine one: $\Delta r_s$, $\Delta s$, and $\Delta \alpha$. For all the plots, the plane is chosen at the $z$ height of the supercell that contains the interstitial. Furthermore, we depict the atoms in this plane with light gold in the third column for guidance. The 4th row shows $\alpha$ with the same range as $s$ to make sure the observed changes in $\alpha$ beyond the next neighbors of the defect are physical and not due to a smaller plot range.}
    \label{fig:2d_ingredients_T_interstitial}
\end{figure*}

\begin{figure}[htb]
  \includegraphics[width=0.89\linewidth]
  {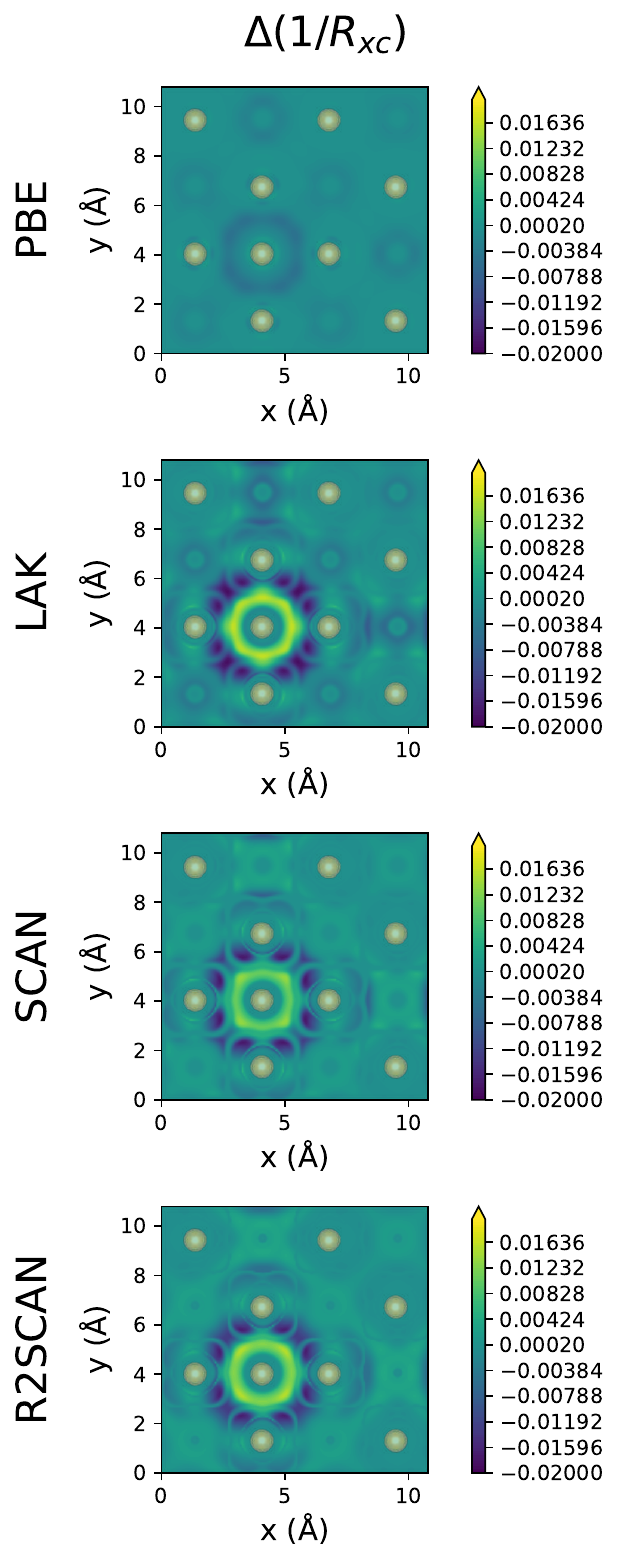}
  \caption{$\Delta (1/R_\mathrm{xc})=(e_\mathrm{xc}^\mathrm{pris}/e_\mathrm{xc}^\mathrm{def})^\mathrm{LDA} -  (e_\mathrm{xc}^\mathrm{pris}/e_\mathrm{xc}^\mathrm{def})^\mathrm{DFA}$ for all DFAs. Atoms are indicated in light gold for guidance. Blue (green) regions have negative (positive) values of $\Delta (1/R_\mathrm{xc})$.}
    \label{fig:delta_Rxc_si-diamond}
\end{figure}

As for the metals, we use unrelaxed structures for the ingredient analysis. For the defect formation energies using the unrelaxed structures, see Fig.~S22 in the Supplemental Material \cite{supplemental}. However, the qualitative trends remain the same as in Fig.~\ref{si-interstitials}.

The analysis for the Si-diamond interstitials is more complex than for the monovacancies in fcc metals. To study the trends of the defect formation energies observed in Fig.~\ref{si-interstitials}, we focus on the T interstitial and identify the important regions with significant changes of the semilocal ingredients in a 2D plane. Figures \ref{fig:2d_diamond_sketch_side} and \ref{fig:2d_diamond_sketch_plane} depict how the 2D region is chosen such that the plane intersects precisely at the height of the interstitial site in both the defective and pristine supercells.

Figure \ref{fig:2d_ingredients_T_interstitial} shows the 2D maps of the DFA ingredients $r_s$, $s$, and $\alpha$, along with the difference between the respective ingredient in the defective and pristine systems ($\Delta$ ingredient) for the T interstitial. Finally, Fig.~\ref{fig:delta_Rxc_si-diamond} shows the 2D maps of the difference $\Delta (1/R_\mathrm{xc})$. Unlike the use of $\Delta R_{xc}$ in the ingredient analysis for fcc metals, for interstitial defects we instead use $\Delta (1/R_\mathrm{xc})$ because the defective system has a higher energy density. Therefore, if we want a metric that lies (almost everywhere) between zero and one, differences of $1/R_\mathrm{xc}$ are more appropriate. We observe the following trends:

\begin{enumerate}
\item[i)] The first row of Fig.~\ref{fig:2d_ingredients_T_interstitial} depicts the regions with large (green) and small (dark blue) $r_s$ for both the pristine and defective systems. These regions correspond to small and large density regions respectively and match with the atomic positions of the supercells. Furthermore, the $\Delta r_s$ frame shows the regions with the most significant changes. We see a noticeable dark blue region right on the interstitial, indicating a drop of $\Delta r_s$ and consequently an increase in the density of the interstitial atom $n^\mathrm{def}(\mathbf{r})$. On the other hand, there are subtle $\Delta r_s$ blue regions between atoms past the first nearest neighbors of the interstitial, suggesting that the perturbation induced by the interstitial extends beyond its immediate vicinity.

\item[ii)] The second row of Fig.~\ref{fig:2d_ingredients_T_interstitial} depicts $s$ for the pristine and defective systems. We see rings of $s$ around each atom characterized by a large $s$ (intense green) followed by a drop to smaller $s$ (dark) as we go further from one atom, resembling the atomic shell structure. Towards the bond centers, $s$ first increases to about one before it decreases to a minimum at the center of each bond. Subsequently, $\Delta s$ shows the regions with the most significant change in $s$. Except for the atomic shell structure, $\Delta s$ becomes negative around the interstitial. As for $\Delta r_s$, there are small regions of slight negative $\Delta s$ near the bond centers past the first neighbors.

\item[iii)] The third and fourth rows of Fig.~\ref{fig:2d_ingredients_T_interstitial} show the trends in $\alpha$, which exhibit significant changes in several regions of the plane. Like $s$, $\alpha$ recognizes the shell structure around the atoms. Whereas $\alpha$ is small near the atoms, it becomes large in the regions furthest from the atoms, as indicated by the intense green regions. Considering the interstitial atom, we see a dark blue region (negative values) of $\Delta \alpha$ followed by a dark red region (positive values) of $\Delta\alpha$. Furthermore, looking at the region around the first nearest neighbors of the interstitial, we see an arrow-shaped red region corresponding to positive values of $\Delta \alpha$ (i.e., $\alpha$ is larger in the defective system and hence there is less enhancement in the defective system, which implies less negative energy and therefore larger defect formation energy). At the tips of the arrow, there are rectangular-shaped dark blue regions, whereas along the edges of the arrows, there are light-blue regions. Interestingly, beyond the blue rectangles in x- and y-direction, there is another red arrow pointing in the opposite direction. Finally, there are slight positive (red) $\Delta \alpha$ regions around the second nearest neighbor atoms of the interstitial. This shows that $\alpha$ captures significant long-range information that is not accessible to $r_s$ and $s$.

\item[iv)] Figure \ref{fig:delta_Rxc_si-diamond} shows the ratio $\Delta(1/R_\mathrm{xc})$ for all DFAs in 2D. Specifically, we observe the following:

\begin{enumerate}
    \item Taking the interstitial atom as a reference and considering first the meta-GGAs, i.e., LAK, SCAN, and $r^2$SCAN, we see a region of positive $\Delta (1/R_\mathrm{xc})$ (green) followed by a region of negative $\Delta (1/R_\mathrm{xc})$ (dark blue). Furthermore, the region close to the first nearest neighbors of the interstitial present an arrow-shaped blue region corresponding to negative values of $\Delta (1/R_\mathrm{xc})$ (and hence positive $\Delta \alpha$), and a small region of negative $\Delta (1/R_\mathrm{xc})$ just below each arrow-shaped region. For LAK, these arrow-shaped areas are larger than for SCAN and $r^2$SCAN. 
    Furthermore, the regions around second-nearest neighbor atoms of the interstitial show a slight negative $\Delta (1/R_\mathrm{xc})$, which matches with regions of slight positive $\Delta \alpha$. Therefore, variations in $\Delta (1/R_\mathrm{xc})$ for the meta-GGAs can be mainly attributed to changes in $\Delta \alpha$, and since the defect formation energies are directly linked to $\Delta (1/R_\mathrm{xc})$, we can dissect the origin of the defect formation energy trends for the T interstitial in Si-diamond.
    
    \item For all three meta-GGAs, $\Delta (1/R_\mathrm{xc})$ exhibits similar features. In particular, several regions beyond the neighboring atoms of the interstitial show significant structure in the meta-GGAs, in contrast to the GGA(PBE), which lacks a dependence on $\alpha$ and whose map of $\Delta (1/R_\mathrm{xc})$ exhibits much less detailed structure than those of the meta-GGAs. Moreover, for the meta-GGAs the map of $\Delta (1/R_\mathrm{xc})$ closely aligns with the features in the map of $\Delta\alpha$, but not with the map of $s$. This shows that $\alpha$ plays a dominant role for the performance of meta-GGAs regarding the defect formation energies of the Si interstitials. These observations together support our finding that $\alpha$ enables meta-GGAs to capture more nonlocal information and that this is crucial for their superior (semiconductor) defect formation energies compared with LDA and GGA.
    
    \item LAK yields even more finely resolved $\Delta (1/R_\mathrm{xc})$ regions than SCAN and r$^2$SCAN. For example, the positive (green) and negative (dark) areas around the interstitials appear more pronounced, and around the bond centers, extending beyond the second nearest neighbors, there are rings of negative $\Delta (1/R_\mathrm{xc})$ that are much less evident in the other meta-GGAs. As a consequence, the defect formation energy of the T interstitial in Si-diamond is even larger with LAK.
The presence of long-ranged features in the meta-GGA maps, particularly for LAK, supports the finding that the orbital dependence not only provides meta-GGAs access to ultranonlocal features of the potential \cite{aschebrock2023exact} but also extends the range of interactions accessible to meta-GGAs qualitatively beyond that of GGAs \cite{lebeda2025meta}. This additional nonlocality is beneficial in semiconductors, but often not desirable in (transition) metals.

\end{enumerate}

\end{enumerate}

\section{Conclusions}

In this work, we studied the formation energies of monovacancies in fcc metals and interstitial defects in semiconducting Si-diamond. We observed opposing trends for semiconductors and the simple metal Al on the one hand and transition metals on the other. For transition metals, all density functionals are reasonably accurate for systems composed of atoms with small atomic radius, and accuracy drops as the mass and atomic radius of the atoms increases. For the transition metals, LDA performs best, LAK performs worst, and the GGA PBE is on average comparable to the advanced meta-GGAs SCAN and $r^2$SCAN.

For semiconductors, the meta-GGAs clearly outperform LDA and PBE, and LAK even outperforms the screened hybrid HSE, often considered the DFT gold-standard for defects in semiconductors. This suggests that LAK can provide a computationally very attractive alternative to HSE for studying defects in semiconductors, and perhaps semiconductor applications more generally.

A key insight of this work comes from the ingredient-level analysis of semilocal functionals. By examining the behavior of $r_s$, $s$, and $\alpha$ along the bond path from atom to vacancy, we identify the spatial regions that dominate defect formation energies and how different ingredients control functional performance. We find that the reduced gradient $s$ is systematically larger in defective systems in the energetically dominant region, which explains the tendency of PBE to stabilize defects relative to LDA. More importantly, the differences among meta-GGAs are governed by their treatment of the iso-orbital indicator $\alpha$. In transition metals, where $\alpha$ exhibits significant variation due to increased orbital overlap and $s$–$d$ hybridization, functionals with stronger $\alpha$ dependence—such as LAK—favor the defective system more strongly, leading to smaller formation energies. In contrast, for simple metals like Al, where the electronic environment is more homogeneous and $\alpha \approx 1$ near the bond center, this effect is reversed. Finally, for the Si interstitial defects we provide direct evidence that $\alpha$ captures significantly nonlocal information, and that this enables meta-GGAs with an enhanced $\alpha$ dependence to achieve excellent performance for semiconductor defects.

Ultimately, we strive for a universal (semilocal) functional that provides a unified description of metals and semiconductors. On the meta-GGA level, adding the Laplacian of the density to discriminate semiconductor and metallic regions is a promising option, as the density Laplacian can provide additional information. A locally scaled self-interaction correction for strongly constrained meta-GGAs could serve a similar purpose. Work in both directions is ongoing \cite{ramasamy2025tackling, shahi2026local}.


\section{Data Availability}
The data and scripts to generate the figures of this work are publicly available \cite{vega_bazantes_2026_21575218}.

\begin{acknowledgments}
This work was supported by the U.S. DOE, Office of Science, Basic Energy Sciences (BES), Grant No. DE-SC0014208. This research used resources of the National Energy Research
Scientific Computing Center, a DOE Office of Science User Facility
supported by the Office of Science of the U.S. Department of Energy
under Contract No. DE-AC02-05CH11231 using NERSC award
BES-ERCAP0036501. The work of JPP was supported by the National Science Foundation under grant nos. DMR-2426275 and CHE-2533416.

\end{acknowledgments}

\appendix


\bibliography{bibliography}
\end{document}


\title{Supplemental Material for ``Rationalizing defect formation energies in metals and semiconductors with semilocal density functionals''}

\author{Jorge Vega Bazantes}
\author{Timo Lebeda}
\email[Contact author: tl681@cam.ac.uk; Current address: Yusuf Hamied Department of Chemistry,
University of Cambridge, Cambridge CB2 1EW, UK]{}
\author{Akilan Ramasamy}
\author{Kanun Pokharel}
\author{Ruiqi Zhang}
\author{John P. Perdew}
\author{Jianwei Sun}\thanks{Contact author: jsun@tulane.edu}

\affiliation{%
Department of Physics and Engineering Physics, Tulane University, New Orleans, Louisiana 70118, USA
}%

\maketitle

\vspace{-\baselineskip}
\section{Calculated defect formation energies and observations about experimental values}

We show the Tables for the fully relaxed and unrelaxed structures for both the Si-diamond interstitials (Tables \ref{tab:si_diamond_table} and \ref{tab:si_diamond_table_unrelaxed}), and the monovacancies in fcc metals (Tables \ref{tab:relaxed_structures} and \ref{tab:exp_structures}). 
The fully relaxed structures are used for performance assessment, whereas the unrelaxed structures serve for ingredient analysis. For the unrelaxed structures, we adopt the experimental lattice constants reported in Refs.~\onlinecite{hao2012lattice,sun2011self}.

\begin{table*}[htb!]
\centering
\caption{The interstitial defect formation energy in Si diamond from the fully relaxed structures in eV.}
\begin{tabular}{c c c c  }
\hline\hline 
    & Tetrahedral (T) & Hexagonal (H) & Split (X) \\
        LDA    & 3.38 & 3.36 & 3.49 \\
        PBE    & 3.76 & 3.60 & 3.55 \\
        SCAN   & 4.56 & 4.30 & 4.21 \\
        $r^2$SCAN & 4.50 & 4.16 & 4.21 \\
        LAK    & 5.33 & 4.79 & 4.88 \\
        HSE06 & 4.81  &  4.34  & 4.21     \\
        DMC \cite{parker2011accuracy}           & 5.1   &  4.7  & 4.4 \\
        DMC \cite{batista2006comparison}        & 5.05  &  5.13 & 4.94 \\
        DMC \cite{leung1999calculations}        & 5.40  &  4.82 & 4.96 \\
        RPA \cite{kaltak2014cubic}              & 4.93  &  4.33 & 4.20 \\
        CCSD \cite{salihbegovic2023formation}   & 7.13  &  5.56 & 5.30  \\
        CCSD(T) \cite{salihbegovic2023formation}& 6.32  &  4.81 & 4.54  \\

\hline
\hline
\end{tabular}
\label{tab:si_diamond_table}
\end{table*}

\begin{table*}[htb!]
\caption{Defect formation energies of fcc metals from the fully relaxed structures. Mean absolute error (MAE), mean absolute deviation per metal (MAD), and standard deviation ($\sigma$) do not consider HSE. All values are in eV. We considered the following experimental defect formation energies:
$^d$\cite{p2_siegel1982positron,smedskjaer1981vacancy},
$^e$\cite{hautojarvi1979positrons},
$^f$\cite{maier_rein_saile_valenta_schaefer_1979,schaefer1987investigation},
$^{g}$\cite{schaefer1987thermal},
$^{h}$\cite{xing2014vacancy,glensk2014breakdown},
$^{j}$\cite{xing2014vacancy},
$^{k}$\cite{p2_siegel1982positron}, and
$^{m}$\cite{glensk2014breakdown,p2_siegel1982positron}.
The HSE values are taken from Ref.~\onlinecite{xing2014vacancy}. For Ni and Pd we considered ferromagnetic (FM) spin arrangements.}
\label{tab:relaxed_structures}

\begin{tabular}{lcccccccccc}
\hline\hline
Solids & LDA & PBE & SCAN & $r^2$SCAN & LAK & HSE~\cite{xing2014vacancy} & Expt. & MAD per metal & $\sigma$ \\
Ni & 1.68 & 1.45 & 1.50 & 1.50 & 1.42 &  & 1.80$^{d}$ & 0.290 & 0.090 \\
Pd & 1.45 & 1.17 & 1.24 & 1.23 & 0.97 & 1.32  & 1.85$^{f}$ & 0.638 & 0.154 \\
Pt & 0.92 & 0.70 & 0.13 & 0.61 & 0.28 & 1.28 & 1.35$^{g}$ & 0.822 & 0.286 \\
Cu & 1.25 & 1.06 & 1.15 & 1.13 & 1.06 & 1.09 & 1.06$^{h}$ & 0.070 & 0.070 \\
Ag & 1.04 & 0.77 & 0.86 & 0.83 & 0.63 & 0.94 & 0.94$^{j}$ & 0.154 & 0.133 \\
Au & 0.62 & 0.39 & 0.33 & 0.42 & 0.18 & 0.72 & 0.94$^{k}$ & 0.552 & 0.142 \\
Al & 0.70 & 0.64 & 0.71 & 0.77 & 0.83 &  & 0.66$^{m}$ & 0.078 & 0.065 \\
Pb & 0.49 & 0.42 & 0.32 & 0.37 & 0.25 &  & 0.58$^{e}$ & 0.210 & 0.082 \\
\hline
MAE per functional & 0.211 & 0.323 & 0.403 & 0.334 & 0.488 &  &  &  &  \\
\hline\hline
\end{tabular}%

\end{table*}



\begin{table*}[htb!]
\caption{Defect formation energies of semilocal DFAs for the fcc metals with unrelaxed structures.
All values in eV.
For Ni and Pd, we considered ferromagnetic (FM) spin arrangements.
}
\label{tab:exp_structures}

\begin{tabular}{lccccc}
\hline\hline
Solids & LDA & PBE & SCAN & $r^2$SCAN & LAK \\
\hline
Ni & 2.03 & 1.47 & 1.76 & 1.68 & 1.35 \\
Pd & 1.78 & 1.07 & 1.32 & 1.23 & 0.59 \\
Pt & 1.44 & 0.68 & 0.82 & 0.86 & 0.18 \\
Cu & 1.51 & 1.04 & 1.40 & 1.27 & 1.04 \\
Ag & 1.25 & 0.64 & 0.88 & 0.79 & 0.26 \\
Au & 0.84 & 0.14 & 0.41 & 0.32 & -0.38 \\
Al & 0.83 & 0.70 & 0.85 & 1.05 & 0.90 \\
Pb & 0.62 & 0.22 & 0.27 & 0.26 & -0.30 \\
\hline\hline
\end{tabular}%
\end{table*}

\begin{table*}[htb!]
\caption{Comparison of defect formation energies calculated using PBE with and without spin--orbit coupling (SOC) for selected fcc metals with unrelaxed structures.
All values in eV.
Due to the high computational cost of SOC calculations, a $3\times3\times3$ $k$-point mesh was used for Pt and Au, and a $4\times4\times4$ $k$-point mesh for Pb.
Identical $k$-point meshes were employed for calculations with and without SOC. Although these calculations are not fully converged with respect to $k$-point sampling, the use of identical meshes allows for a more consistent assessment of the SOC contribution to the defect formation energies.
}
\label{tab:soc_comparison}

\begin{tabular}{lcc}
\hline\hline
Solids & PBE & PBE+SOC \\
\hline
Pt & 0.53 & 0.59 \\
Au & 0.16 & 0.13 \\
Pb & 0.31 & 0.28 \\
\hline\hline
\end{tabular}%
\end{table*}

\FloatBarrier

We note that most experimental monovacancy formation energies used in this work are taken from Ref.~\onlinecite{erhart1991atomic}, a widely cited compilation in the literature. Where possible, we also consulted the primary references therein to verify the reported values. Specific comments about each metal's experimental data are described below:

\begin{itemize}

    \item \textbf{Ni}: Value appears in Ref.~\onlinecite{erhart1991atomic} as (1.79 $\pm$ 0.05) eV, and is given as the recommended value for single vacancies. Reference \onlinecite{erhart1991atomic} refers to (a preprint of) Ref.~\onlinecite{p2_siegel1982positron} for the original value, which lists (1.8 $\pm$ 0.1) eV. This value is further confirmed in Ref.~\onlinecite{smedskjaer1981vacancy}, which published a value of (1.8 $\pm$ 0.1) eV and hence we use this value. Furthermore, it is worth mentioning recent work in Ref.~\onlinecite{gong2018temperature}, which has a compilation of the experimental values of the formation energy of the monovacancy of Ni. Also, Ref.~\onlinecite{gong2018temperature} performed calculations for the Gibbs energy of vacancy formation in nickel at $T>0$ K.

    \item \textbf{Pd}: There is no recommended value for Pd. Reference \cite{erhart1991atomic} cites Ref.~\onlinecite{hu1979positron} and references therein with values of 1.70 eV and 1.85 eV. The latter value, (1.85 $\pm$ 0.25) eV, is supported by Refs.~\onlinecite{maier_rein_saile_valenta_schaefer_1979,schaefer1987investigation}, so we use 1.85 eV.
    

    \item \textbf{Pt}: Reference \cite{erhart1991atomic} recommends a value of (1.35 $\pm$ 0.05) eV for single vacancies. This value appears in Ref.~\onlinecite{schaefer1987thermal} as well with an uncertainty of 0.09 eV. We use 1.35 eV in our work.
    
    \item \textbf{Cu}: (1.28 $\pm$ 0.05) eV is the recommended value in Ref.~\onlinecite{erhart1991atomic} using all data. Reference \onlinecite{erhart1991atomic} refers to Refs.~\onlinecite{siegel1982point} and \cite{schaefer1987thermal_ref1}. The latter provides a value of (1.28 $\pm$ 0.07) eV, however, the `recommended' value by the same reference is (1.13 $\pm$ 0.04) eV. Furthermore, the more recent Refs.~\onlinecite{xing2014vacancy, glensk2014breakdown} have revisited the experimental values of the Cu monovacancy and suggest using 1.06 eV, which is the value we use in this work.

    \item \textbf{Ag}: As stated in Ref.~\onlinecite{erhart1991atomic}: (1.11 $\pm$ 0.05) eV, is the recommended value for single vacancies at medium temperature, and cites Ref.~\onlinecite{p2_siegel1982positron}. On the other hand, Ref.~\onlinecite{xing2014vacancy} re-assessed the experimental value of the Ag monovacancy formation energy and proposed the value of 0.94 eV, which is the one we use in this paper.
    
    \item \textbf{Au}: (0.93 $\pm$ 0.04) eV is the recommended value for single vacancies at medium temperature by Ref.~\onlinecite{erhart1991atomic}, which cites a preprint of Ref.~\onlinecite{p2_siegel1982positron}. The latter lists a `best' value of (0.94 $\pm$ 0.02) eV, which is the one we use.

    \item \textbf{Al}: As stated in Ref.~\onlinecite{erhart1991atomic}: Recommended value for single vacancies at medium temperature is (0.67 $\pm$ 0.03) eV. References supporting this value are: \cite{balluffi1978vacancy, siegel1978vacancy,schaefer1987positron}. A revised value of 0.66 eV was published in Ref.~\onlinecite{glensk2014breakdown} (which also is considered as best value in Ref.~\onlinecite{p2_siegel1982positron}) although it only differs by 0.01 eV from the already recommended one. Hence we use the value of 0.66 eV.

    \item \textbf{Pb}: Ref.~\onlinecite{erhart1991atomic} recommends a value of (0.58 $\pm$ 0.04) eV and refers to Refs.~\onlinecite{p2_siegel1982positron,siegel1982point}. This value also appears in Refs.~\onlinecite{hautojarvi1979positrons,sharma1976high}, which along with other values presents 0.58 eV, the value recommended in Ref.~\onlinecite{erhart1991atomic}.

    
\end{itemize}

\section{Detailed figures}
We obtain $r_s$ and $s$ from the reconstructed all-electron density provided in \textsc{vasp} (AECCAR0+AECCAR2 files), while we calculate $\alpha$ from the electron localization function of the valence electrons (ELFCAR file). Consequently, in the pseudopotential (core) region close to the nucleus, the values of $\alpha$ are meaningless, whereas $r_s$ and $s$ are reasonable approximations (although $s$ shows sharp peaks there due to the discretization of the grid.).
Below we show the Figures for all the metals depicting the pure ingredients and ratios $R_{xc}$ and their differences $\Delta$.

\metalfig{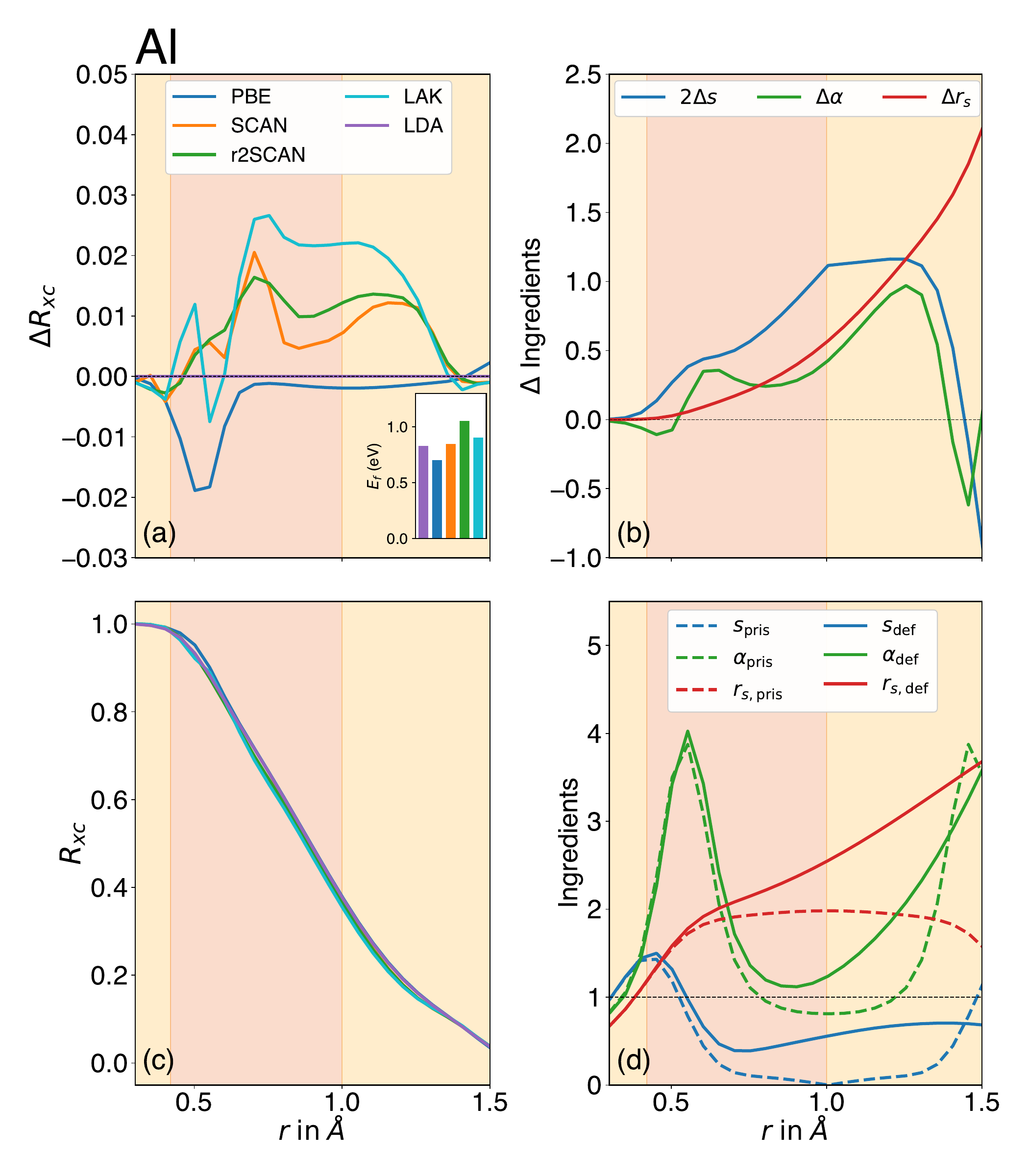}{Al}
\metalfig{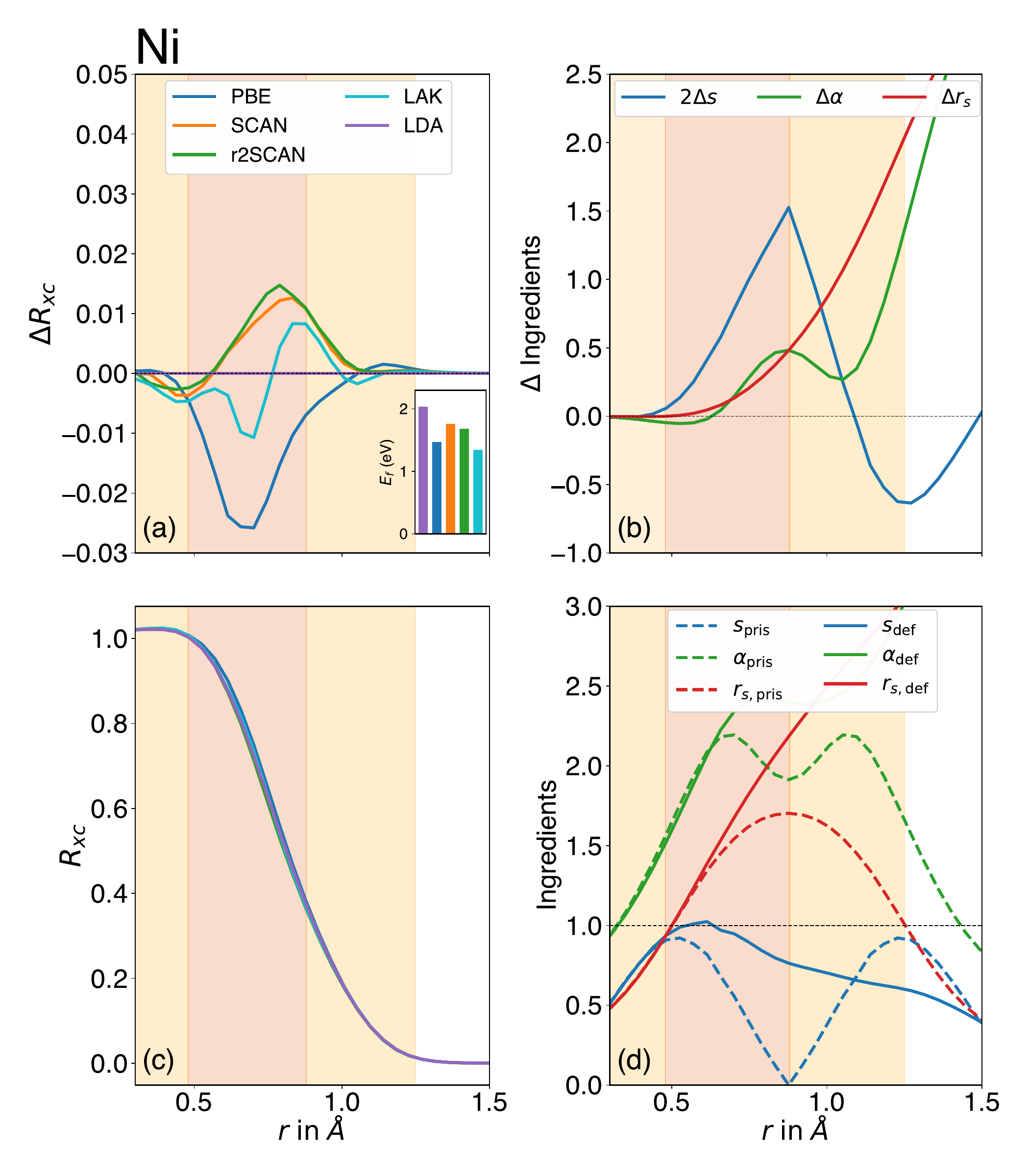}{Ni}
\metalfig{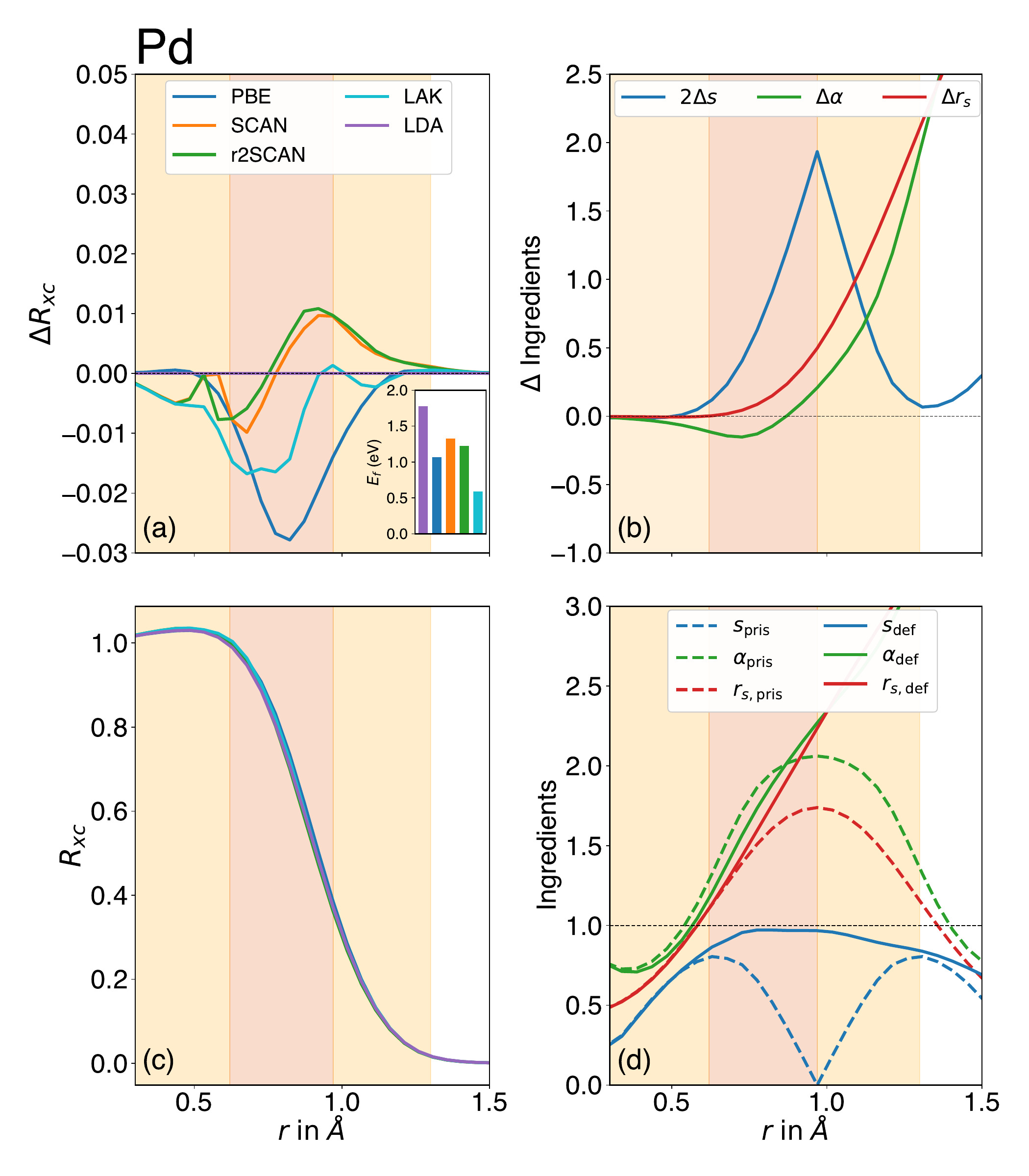}{Pd}
\metalfig{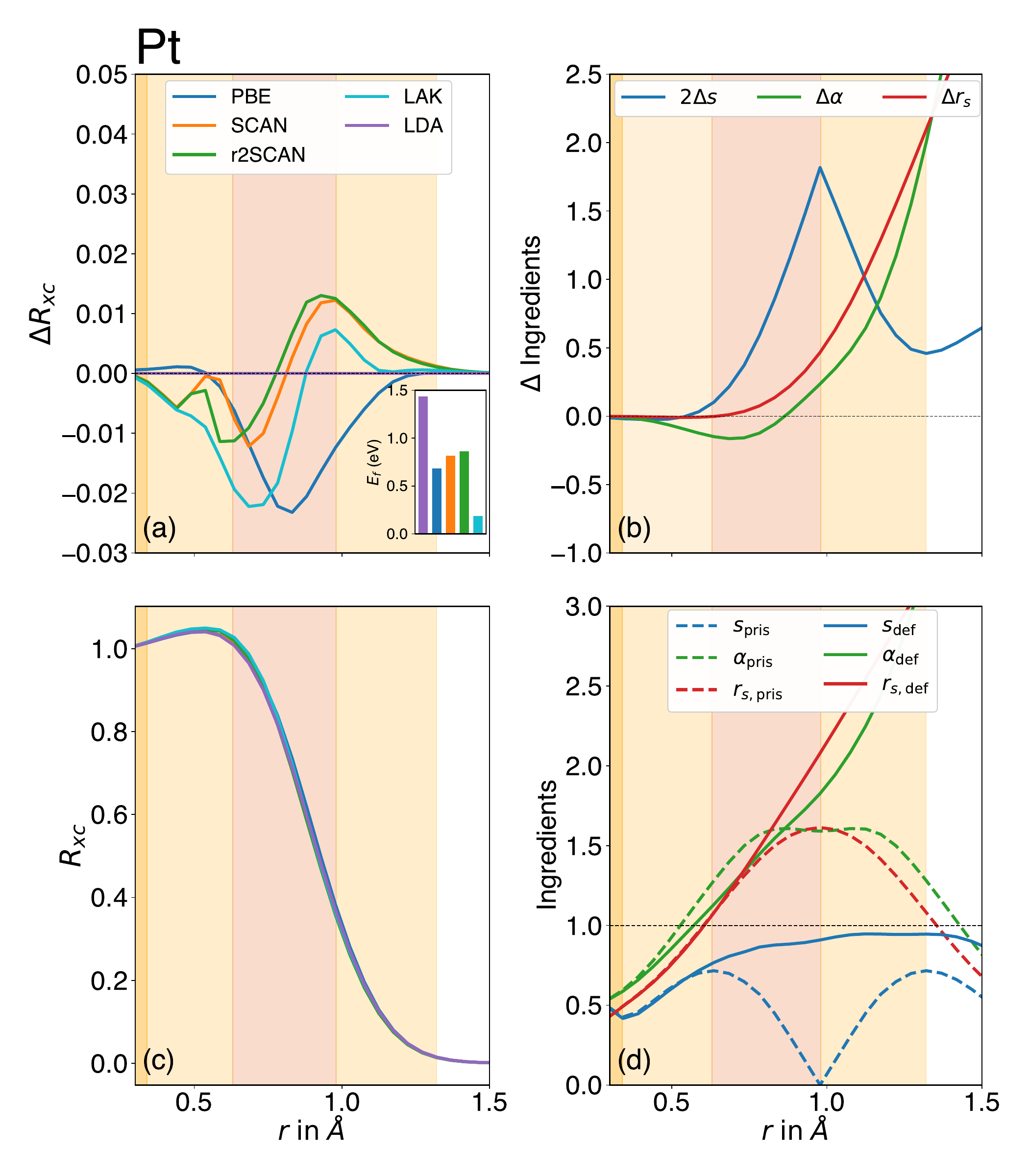}{Pt}

\metalfig{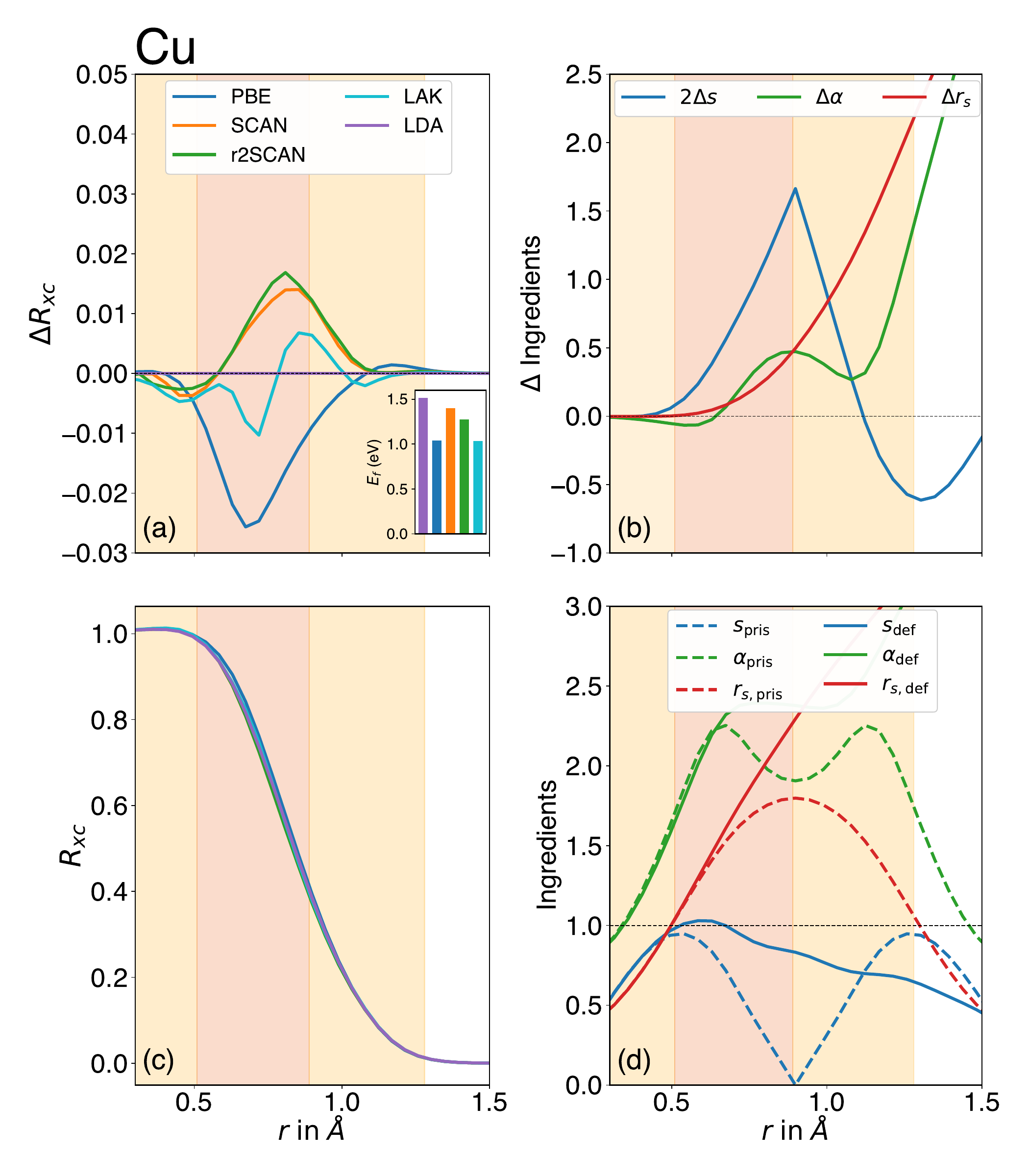}{Cu}
\metalfig{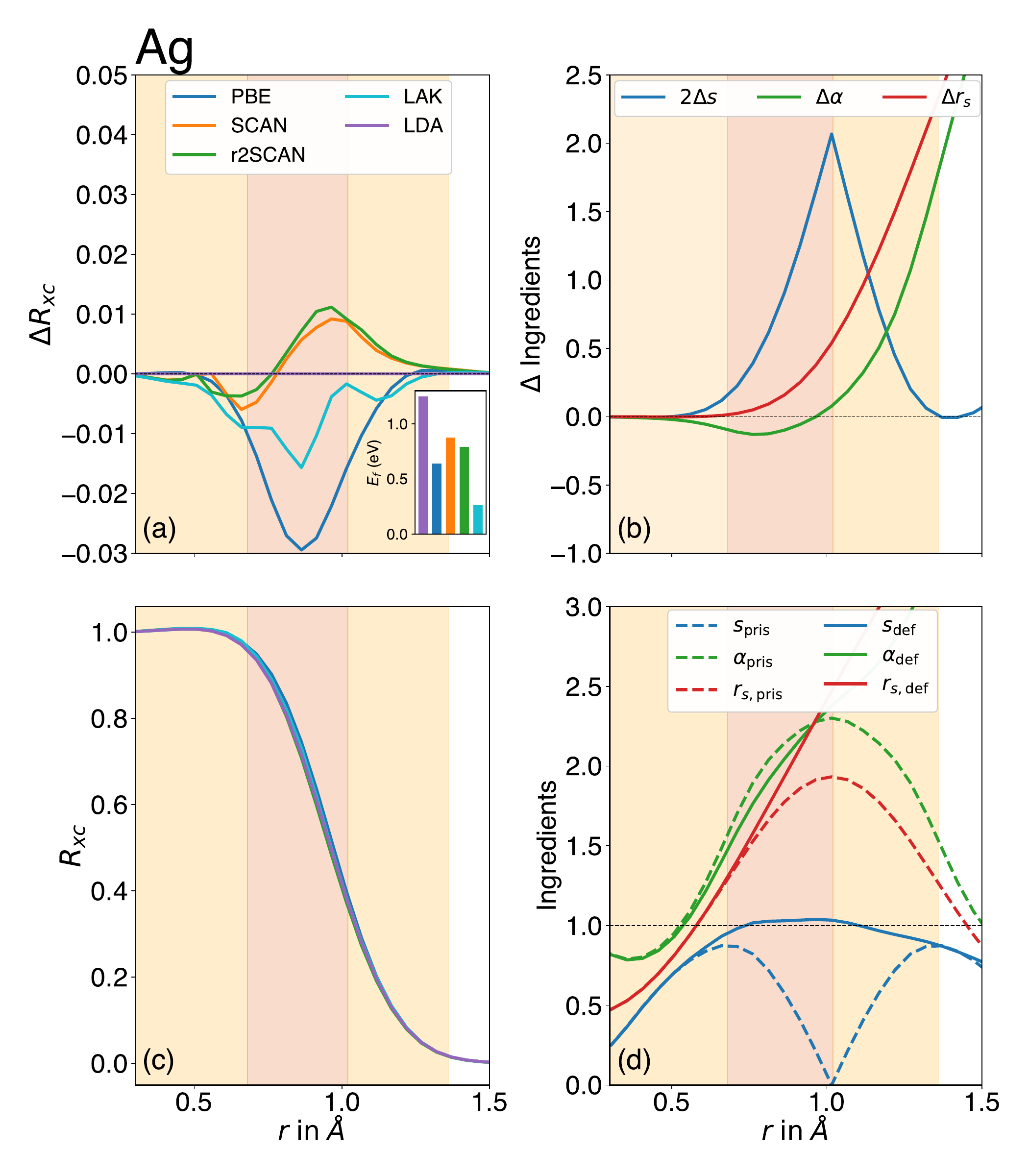}{Ag}
\metalfig{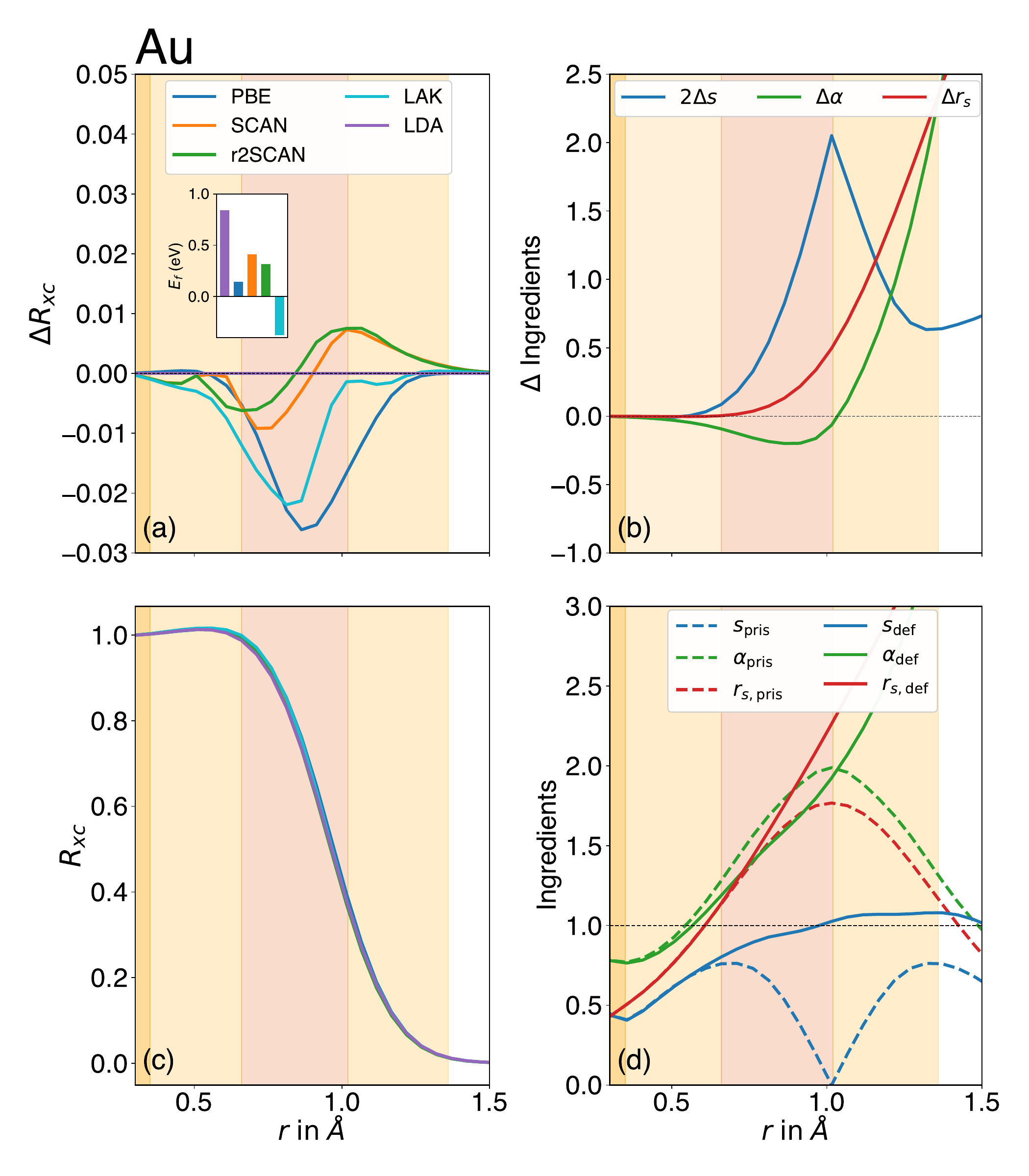}{Au}
\metalfig{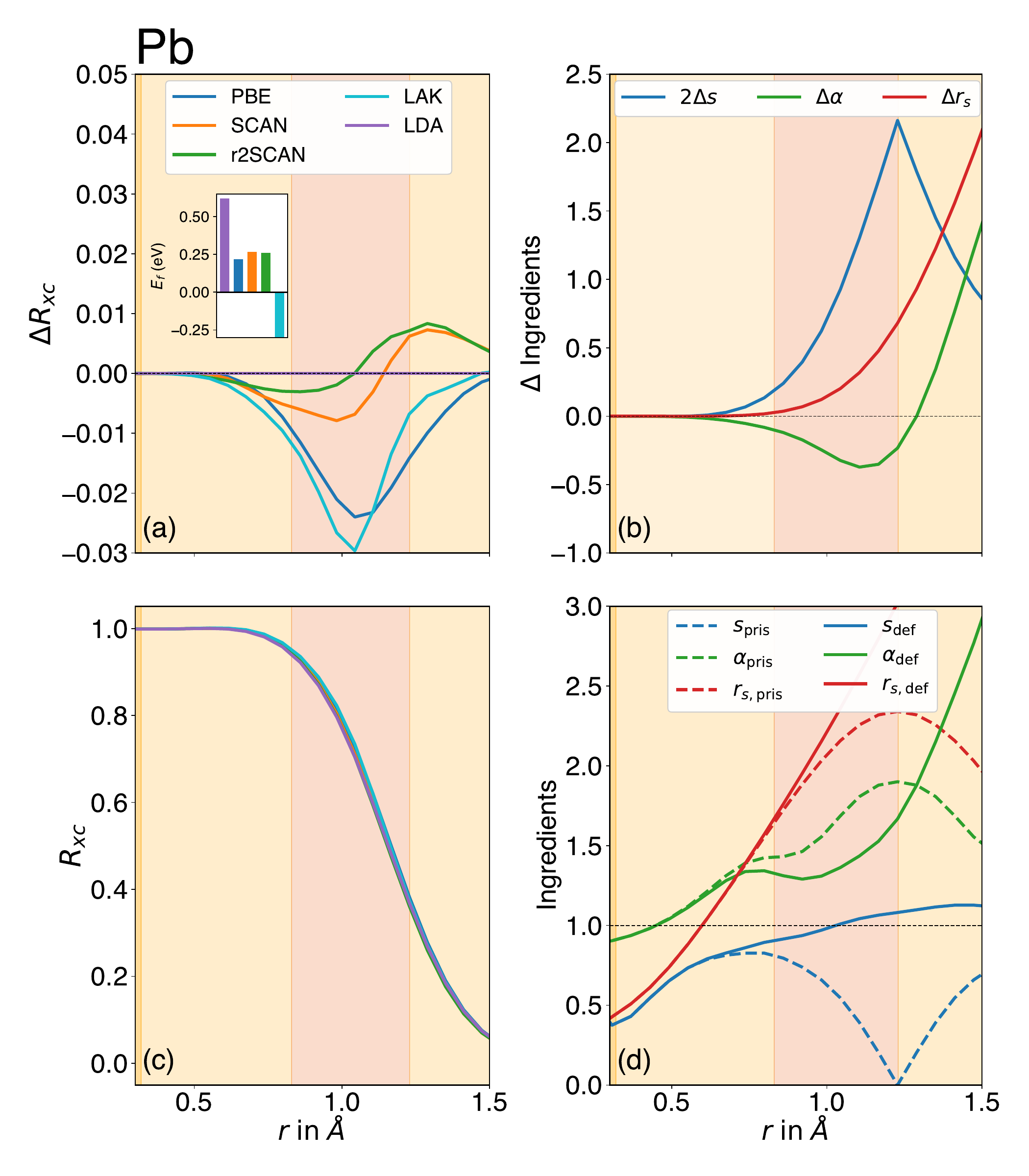}{Pb}


\pagebreak

\subsection{Ingredients of fcc metals: supercells and atoms, the case of Pt}
\begin{figure*}[htb!]
\centering
  \includegraphics[width=\textwidth]{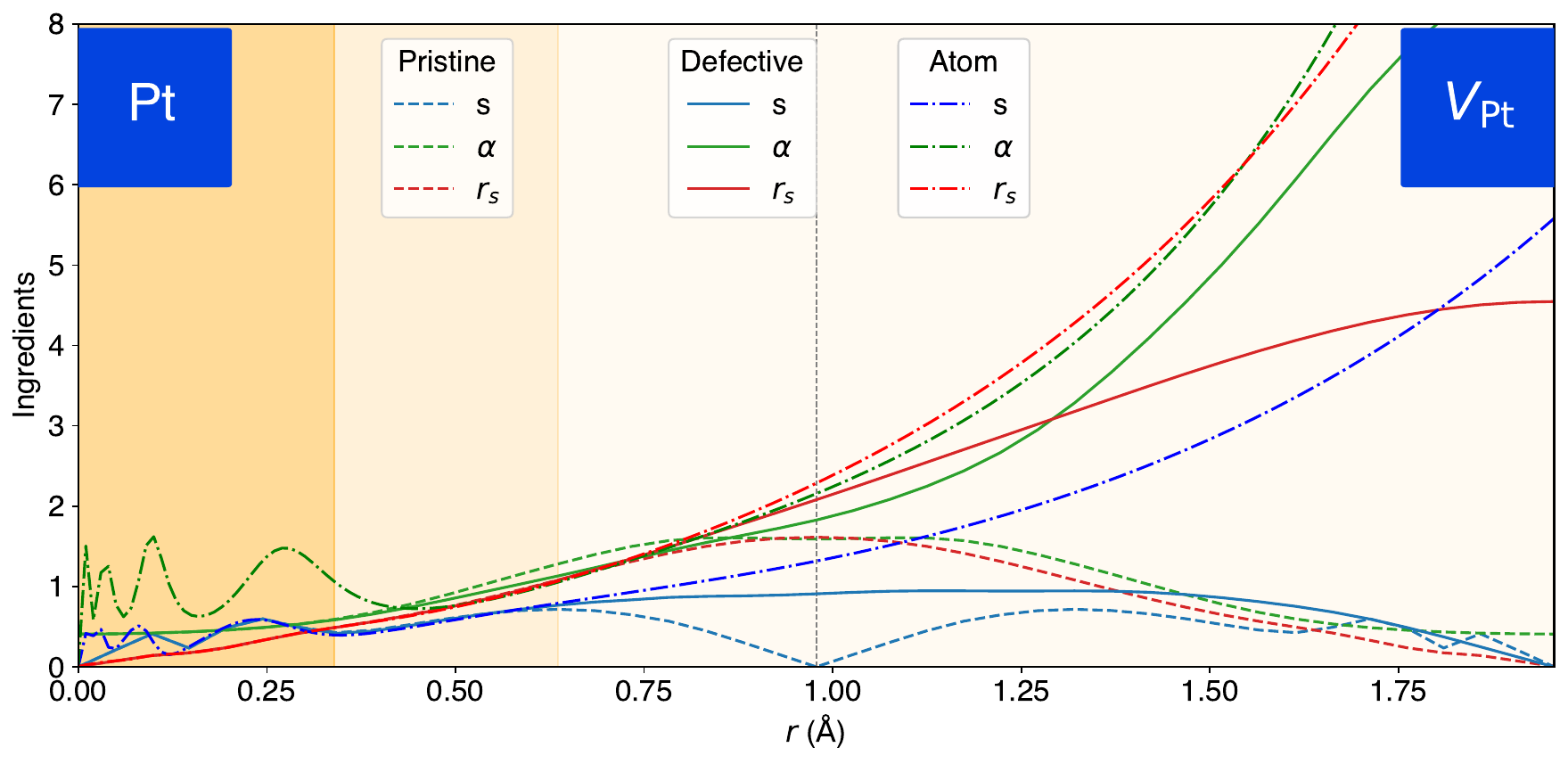}
	\caption{Ingredients analysis of Pt including the atom's. We depict the values of the Wigner-Seitz radius $r_s$, the reduced density gradient $s$, and the iso-orbital indicator $\alpha$ along a path in the unit cell for the pristine and defective systems. Dashed is for the pristine system, lines are for the system with defect, and dash-dot is for the atom. The defective and pristine system values are calculated self-consistently using PBE in \textsc{vasp}, whereas the atom was obtained with \textsc{band}. Note that for all the plots the path is chosen as the shortest path from an atom to the vacancy; one nucleus is on the left and on the right is the second nucleus or the vacancy, respectively. Unrelaxed (experimental) lattice constants are used. The dominant region (pink) spans from the bond center (where $s=0$) to the next maximum in $s$ on the left hand side. In addition, there are two important regions (light orange). One spans from the bond center to the maximum of $s$ on the right hand side, and the other spans from the maximum of $s$ on the left hand side to the next minimum of $s$ on the left hand side. Further, there are unimportant regions. One is the core region (orange), which spans from the latter minimum in $s$ to the center of the nucleus. The other is the vacancy region (white), where the density in the defective system becomes very small.
    }
	\label{fig:ingredients_with_atom}
\end{figure*}

In order to verify the reliability of our supercell ingredients results, in figure \ref{fig:ingredients_with_atom} we depict $\alpha$, $s$, and $r_s$ for Pt computed by \textsc{vasp}, and the ingredients of Pt in a larger box computed with the all electron (AE) code \textsc{band} \cite{band1,band2,band3,band4,band2023.1}. If we look closely to $r_s$ in the core and left important regions, the \textsc{vasp} supercell results match closely with the AE \textsc{band} results for the atom. As we go through the dominant and right important region, $r_s$ for both the defective system and the atom starts to increase, with the atom having a larger and steeper increase. 

On the other hand, regarding the ingredient $s$ for the atom in the core region, the \textsc{vasp} supercell results are a decent approximation although as we go closer to the nucleus the results given by \textsc{vasp} have no longer a good resolution. In contrast, the \textsc{band} result clearly shows node cores corresponding to the s states. Furthermore, $\alpha$ in the core region with the AE calculation depicts clearly the core states, while the \textsc{vasp} results show no nodes. This matches our previous observation that the $\alpha$ values from \textsc{vasp} in the core region are meaningless since $\alpha$ was calculated from the electron localization function of the valence electrons. In the left important region, all $\alpha$ values still differ, and $\alpha$ from the atom only matches in one point in the core region and later in a small section at the beginning of the dominant region. 

\subsection{Figures in 2D of the fcc metals} 
\begin{figure*}[htb!]
    \centering
    \includegraphics[width=0.9\textwidth]{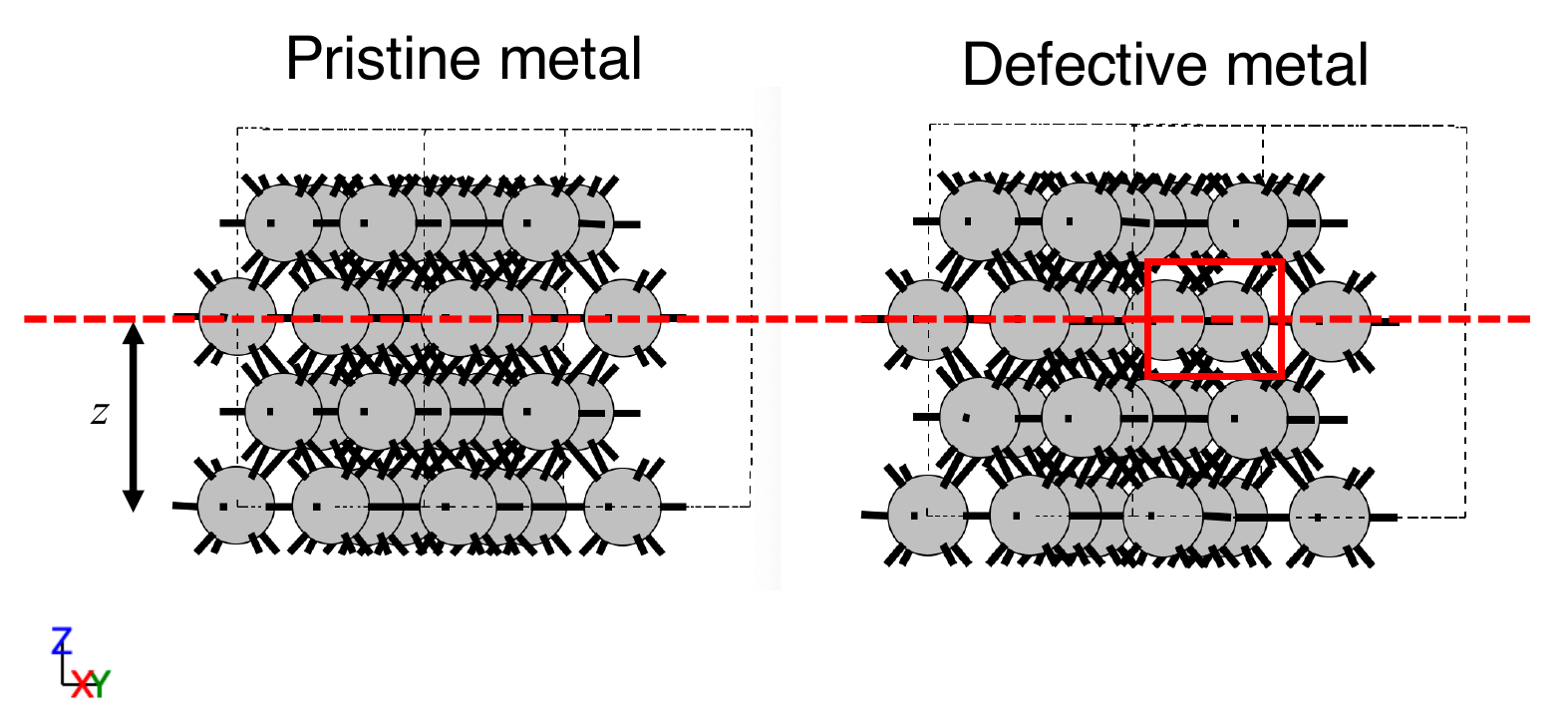}
 \caption{Side view of the pristine (left) and defective system (right) of an fcc metal. The box in red encloses the region with the defect (a vacancy), the red dashed line is the plane at which the cut for the ingredient analysis in 2D is made.}  
 \label{fig:sketch_metals}
\end{figure*}

\begin{figure*}[htb!]
    \centering
    \includegraphics[width=0.9\textwidth]{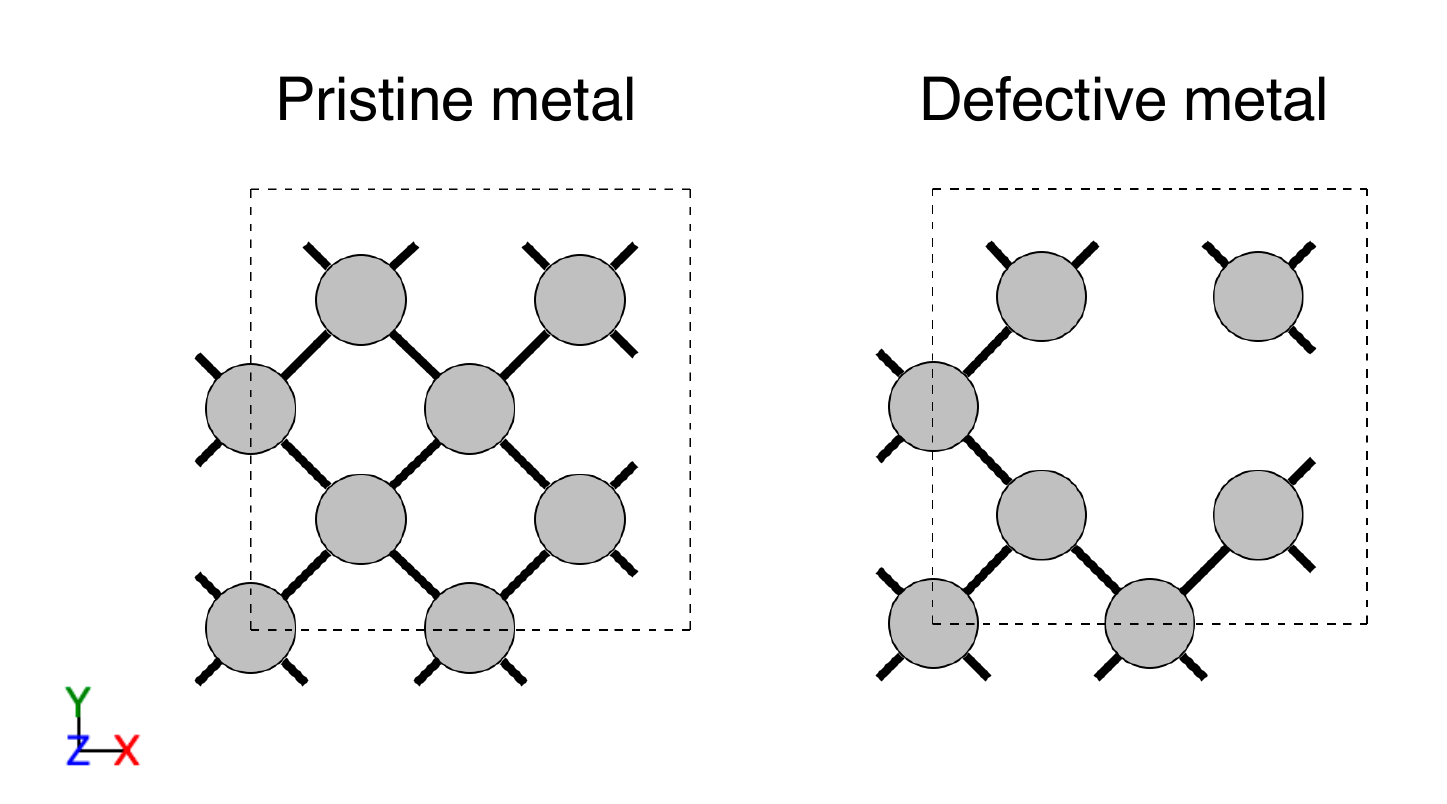}
 \caption{Plane view of the pristine (left) and defective system (right) of an fcc metal. The 2D ingredients plots depict this plane. Atoms of other planes are not shown.}  
 \label{fig:sketch_metals_xy}
\end{figure*}

In this section we show the sketch of the plane for the 2D plots in Figures \ref{fig:sketch_metals} and \ref{fig:sketch_metals_xy}, and present the results of the DFAs ingredients and $R_{xc}$ for all fcc metals in Figures S12$-$S21.

\begin{figure*}[htb!]
\centering
  \includegraphics[width=\textwidth]{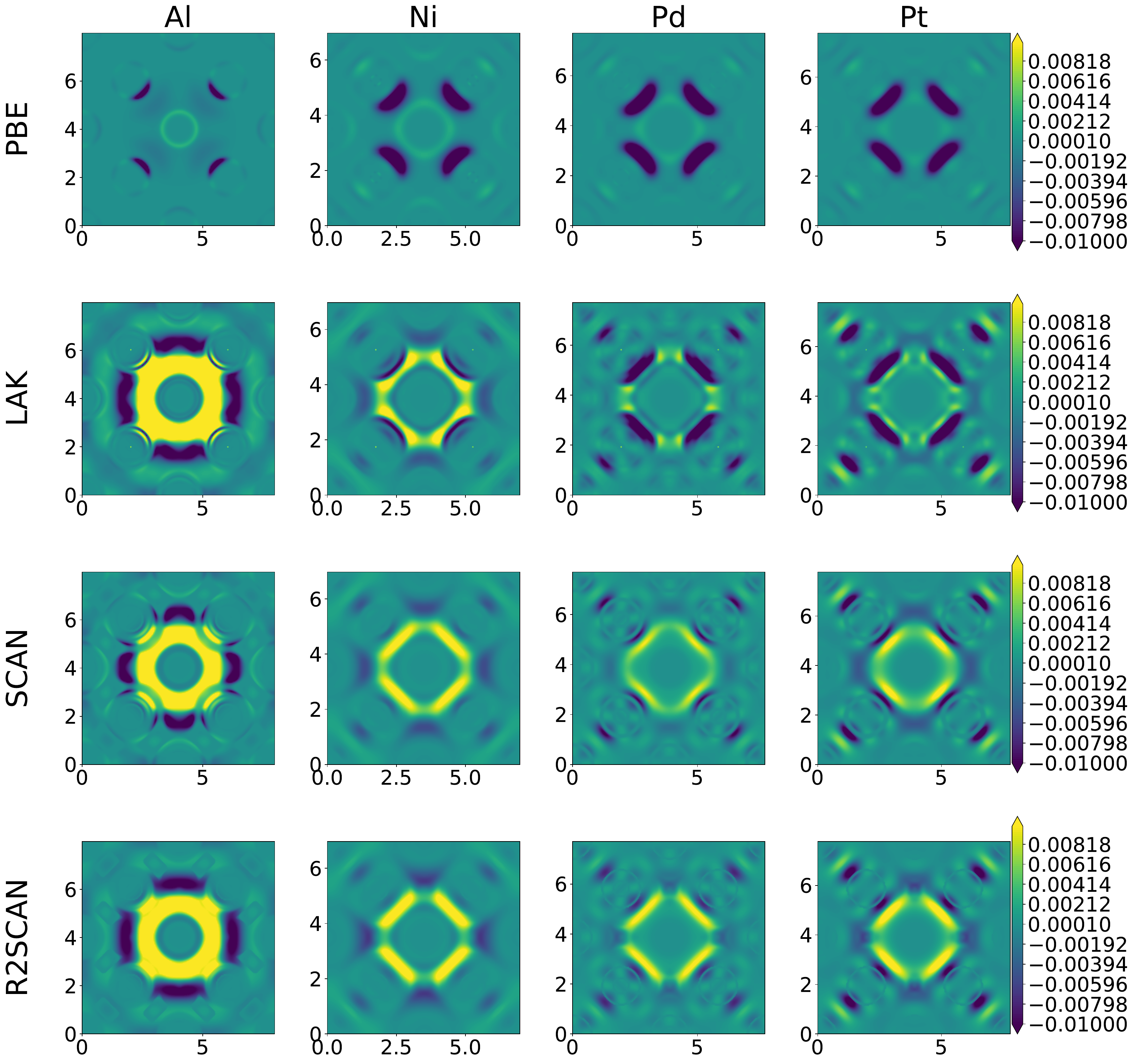}
	\caption{$\Delta R_\mathrm{xc}^\mathrm{DFA} = R_\mathrm{xc}^\mathrm{LDA} - R_\mathrm{xc}^\mathrm{DFA}$ for Al, Ni, Pd, and Pt in the 1st, 2nd, 3rd, and 4th columns respectively. Each row from top to bottom corresponds to the DFAs PBE, LAK, SCAN and $r^2$SCAN, respectively. The vacancy is in the middle of the frame plot. The regions with positive (negative) $\Delta R_\mathrm{xc}^\mathrm{DFA}$ are depicted in green (dark blue), corresponding to $R_\mathrm{xc}^\mathrm{LDA}(\textbf r)>R_\mathrm{xc}^\mathrm{DFA}(\textbf r)$ ($R_\mathrm{xc}^\mathrm{LDA}(\textbf r)<R_\mathrm{xc}^\mathrm{DFA}(\textbf r)$). Positive $\Delta R_\mathrm{xc}^\mathrm{DFA}(\textbf r)$ corresponds to the DFA (locally) favoring the pristine system over the defective one more than LDA does, and thus a contribution to a larger defect formation energy than that of LDA.}
	\label{fig:2d_ingredients_main}
\end{figure*}

\begin{figure*}[htb!]
\centering
  \includegraphics[width=\textwidth]{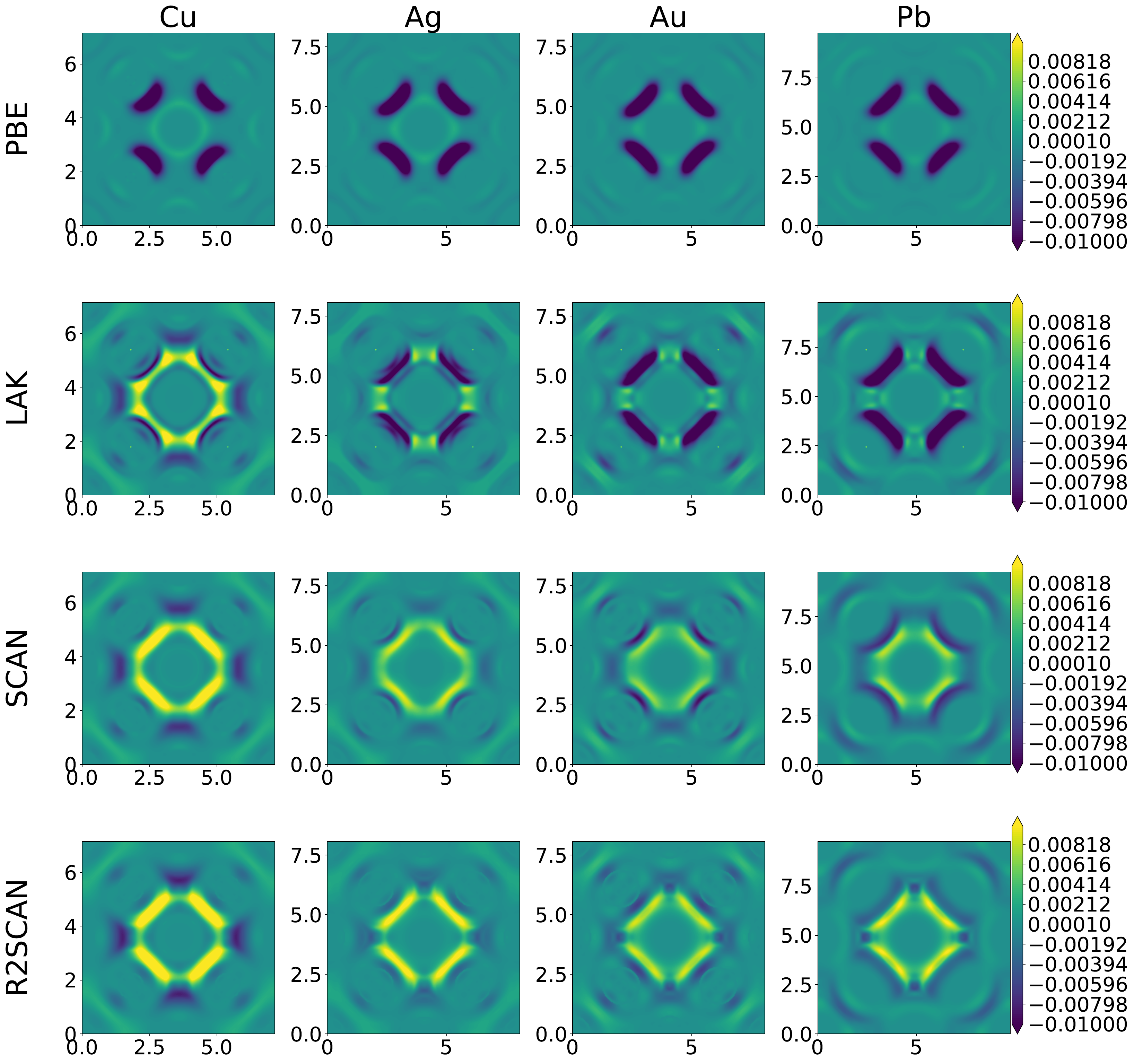}
	\caption{$\Delta R_\mathrm{xc}^\mathrm{DFA} = R_\mathrm{xc}^\mathrm{LDA} - R_\mathrm{xc}^\mathrm{DFA}$ for Cu, Ag, Au, and Pb in the 1st, 2nd, 3rd, and 4th columns respectively. Each row from top to bottom corresponds to the DFAs PBE, LAK, SCAN and $r^2$SCAN, respectively. The vacancy is in the middle of the frame plot. The regions with positive (negative) $\Delta R_\mathrm{xc}$ are depicted in green (dark blue), corresponding to $R_\mathrm{xc}^\mathrm{LDA}(\textbf r)>R_\mathrm{xc}^\mathrm{DFA}(\textbf r)$ ($R_\mathrm{xc}^\mathrm{LDA}(\textbf r)<R_\mathrm{xc}^\mathrm{DFA}(\textbf r)$). Positive $\Delta R_\mathrm{xc}^\mathrm{DFA}(\textbf r)$ corresponds to the DFA (locally) favoring the pristine system over the defective one more than LDA does, and thus a contribution to a larger defect formation energy than that of LDA.}
	\label{fig:2d_ingredients_main}
\end{figure*}


\twodmetalfig{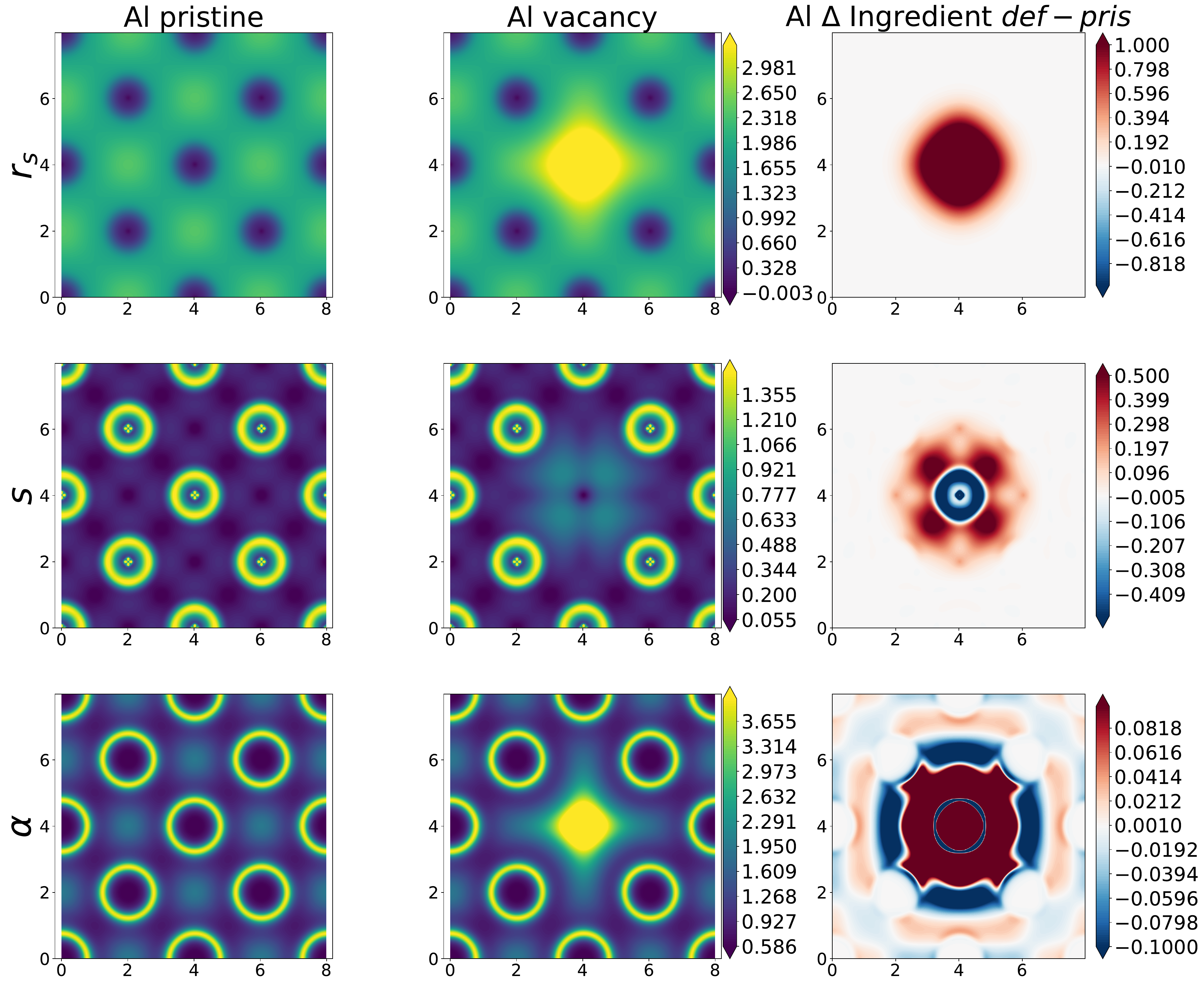}{Al}
\twodmetalfig{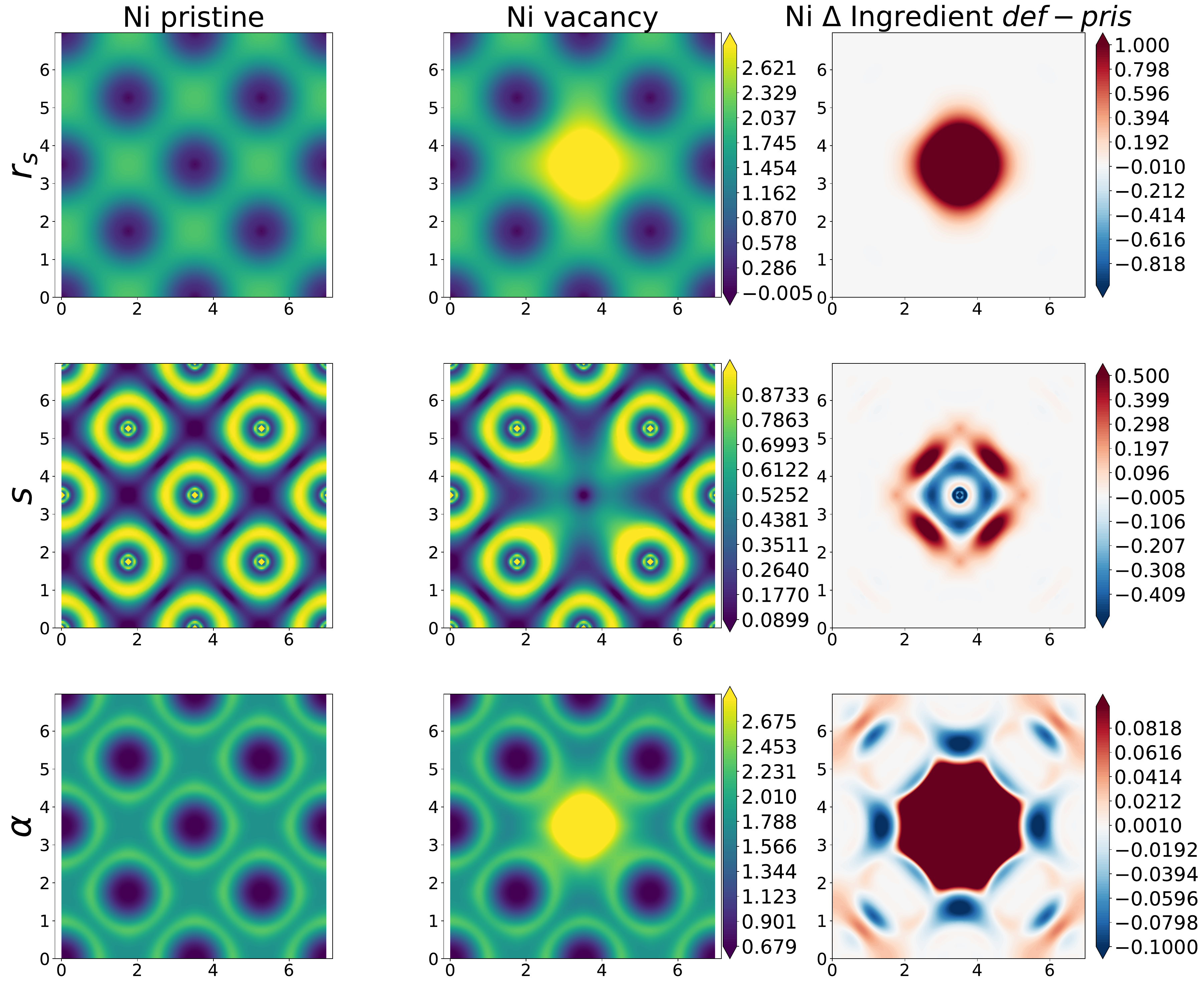}{Ni}
\twodmetalfig{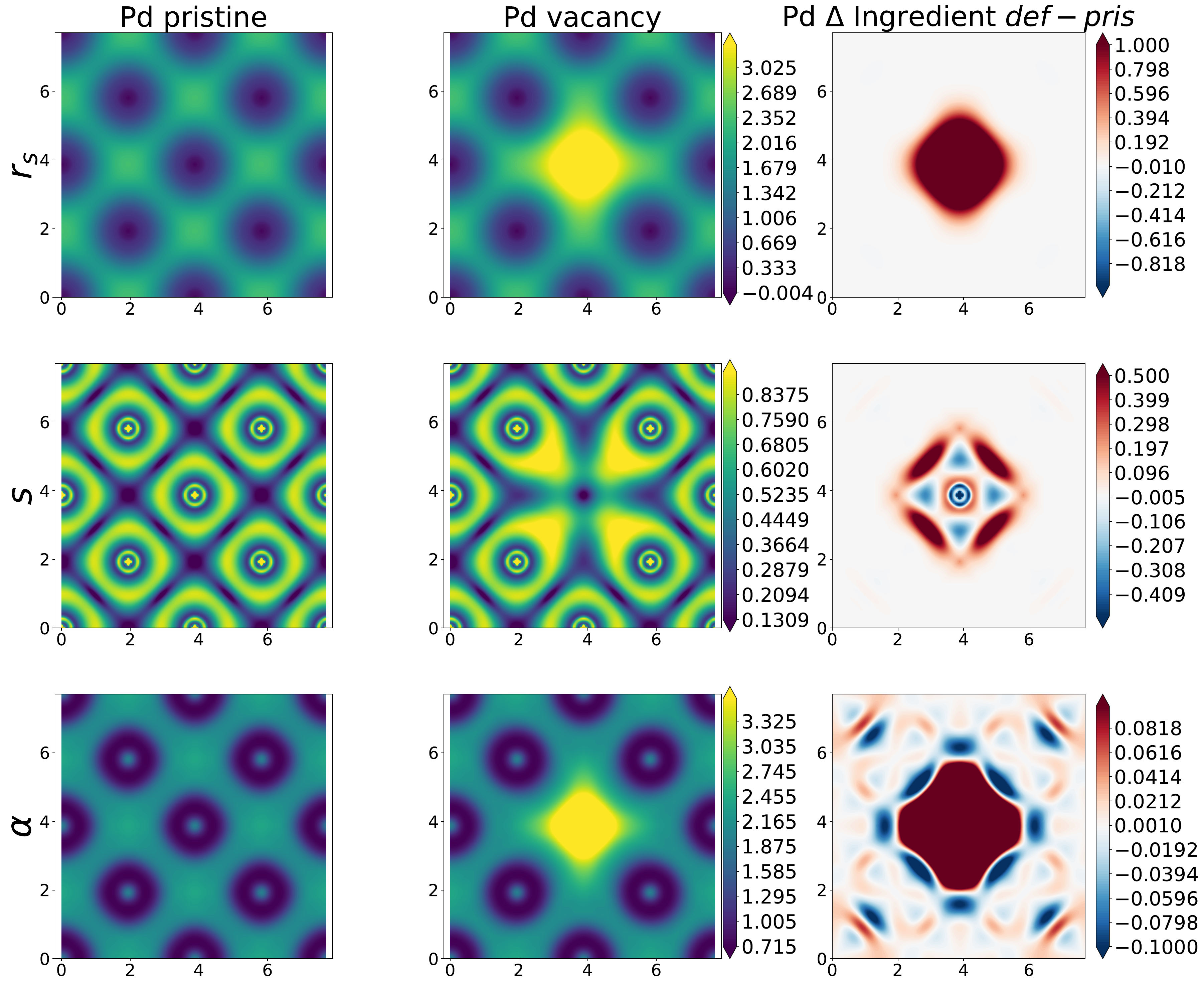}{Pd}
\twodmetalfig{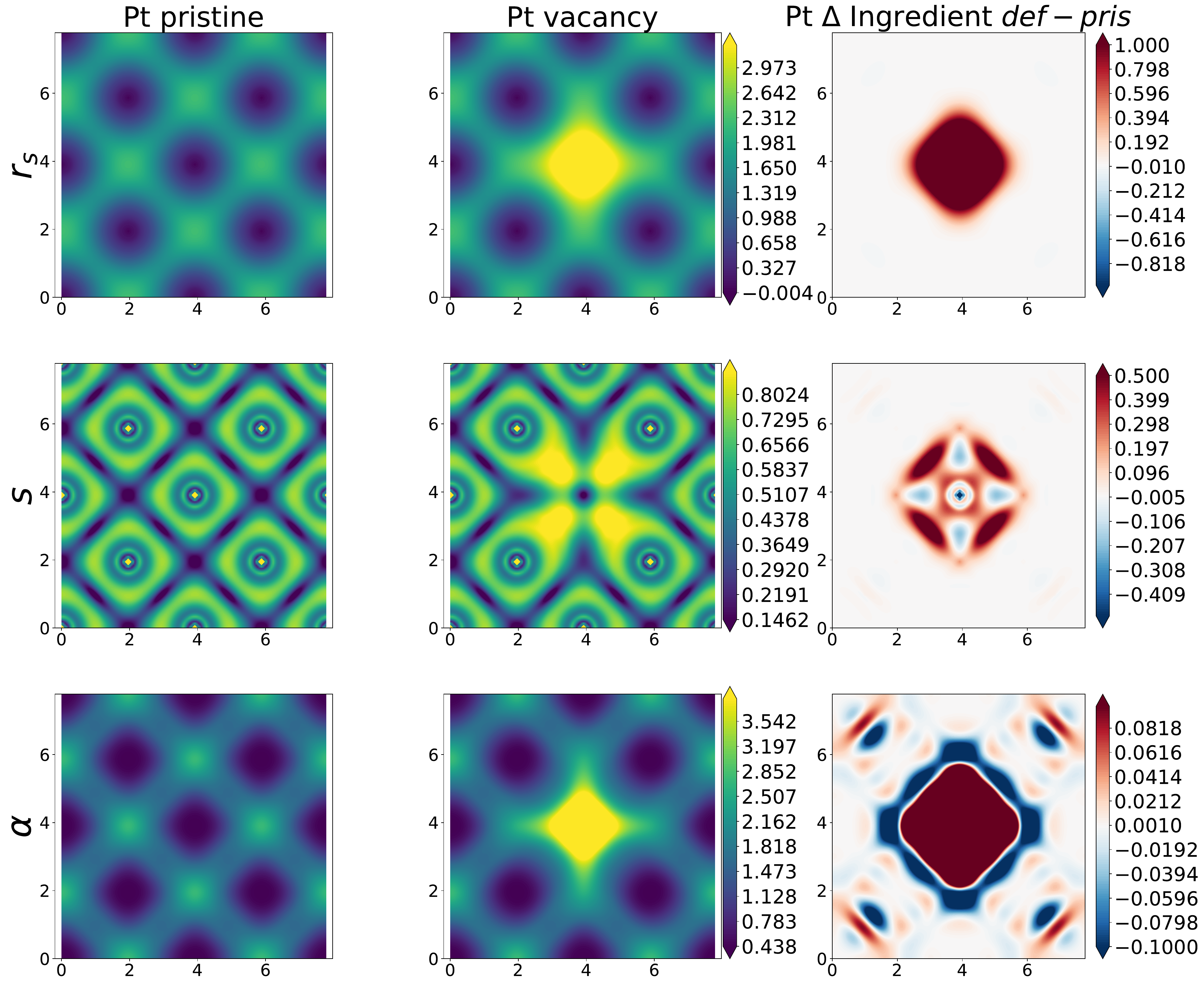}{Pt}

\twodmetalfig{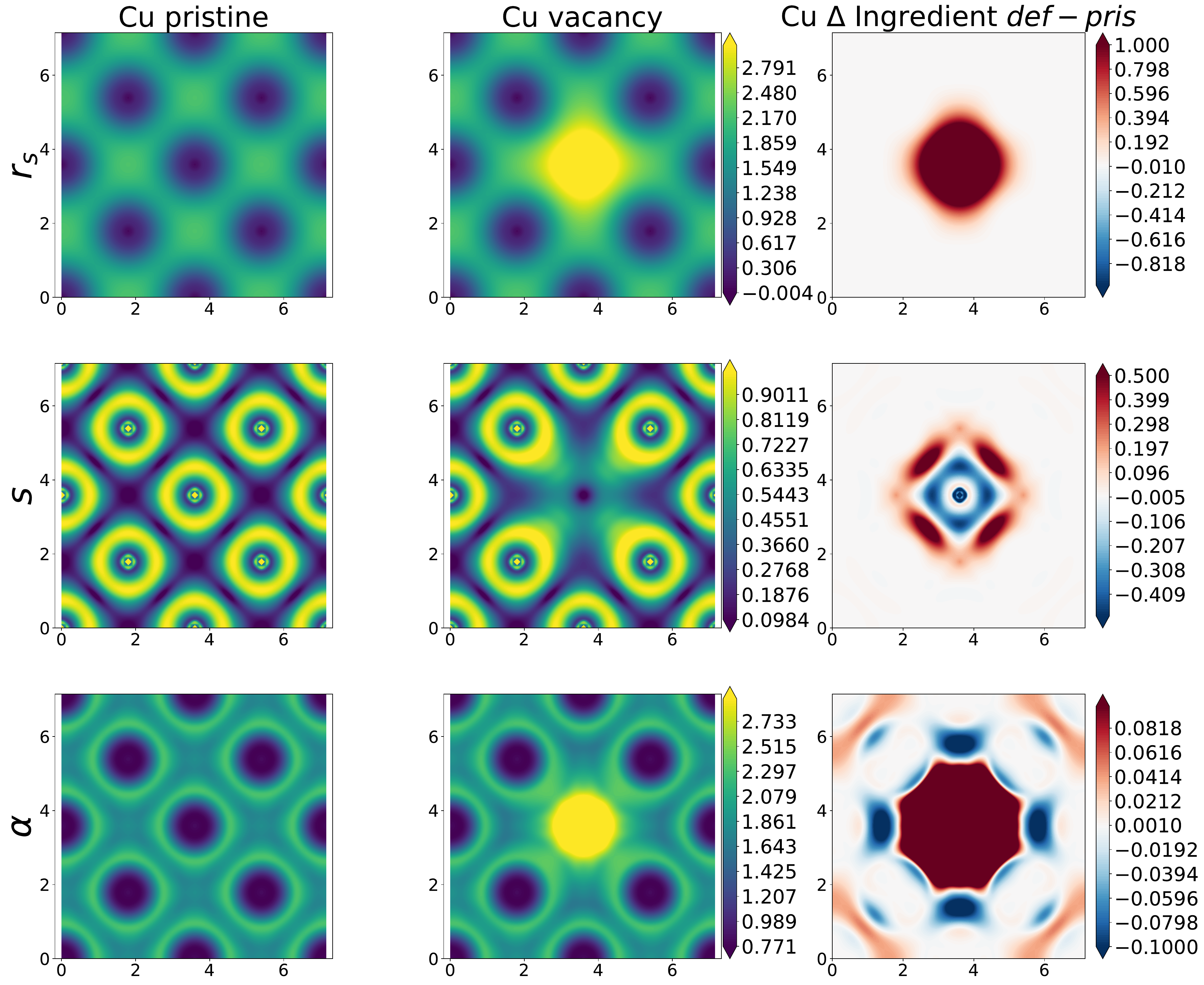}{Cu}
\twodmetalfig{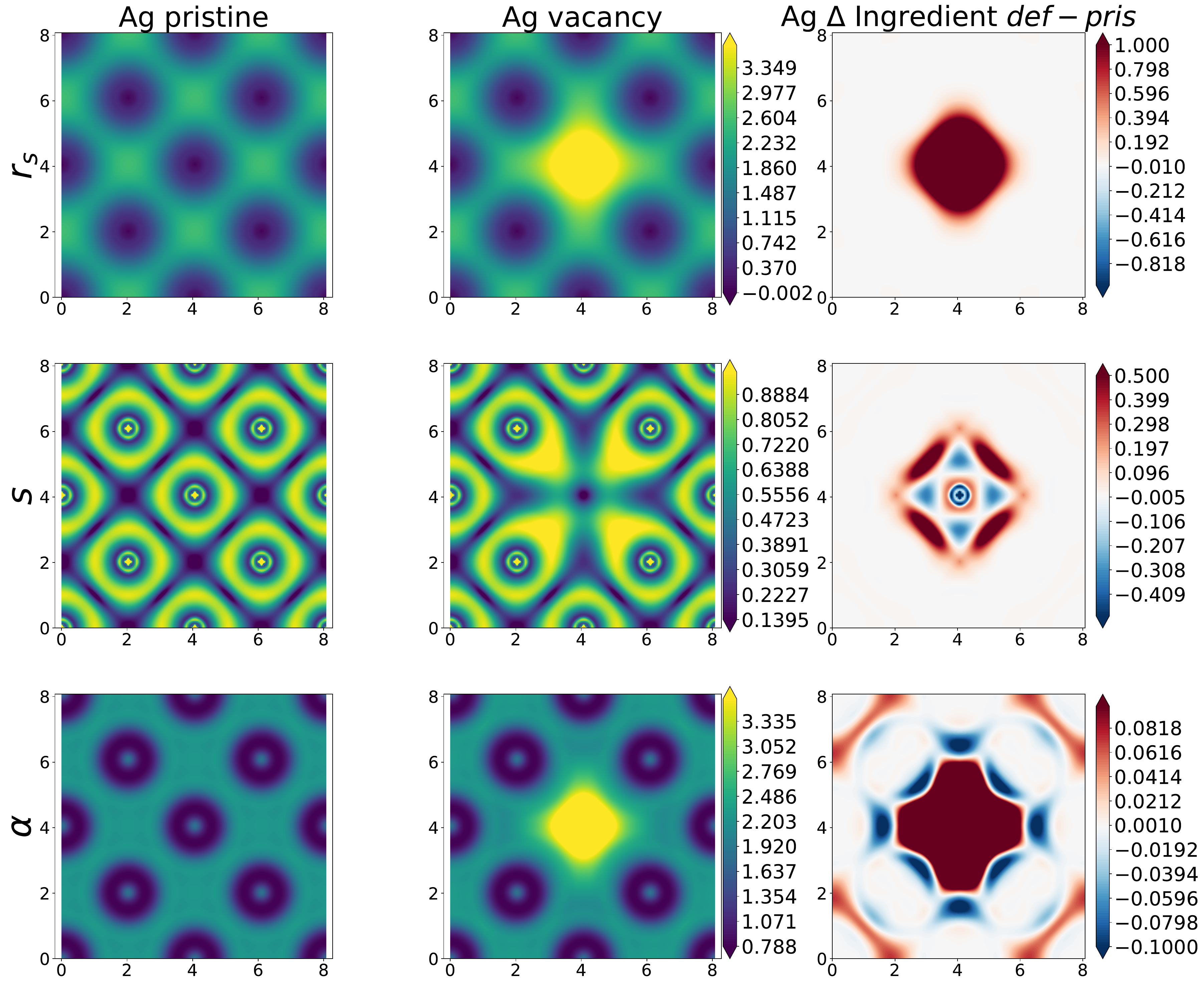}{Ag}
\twodmetalfig{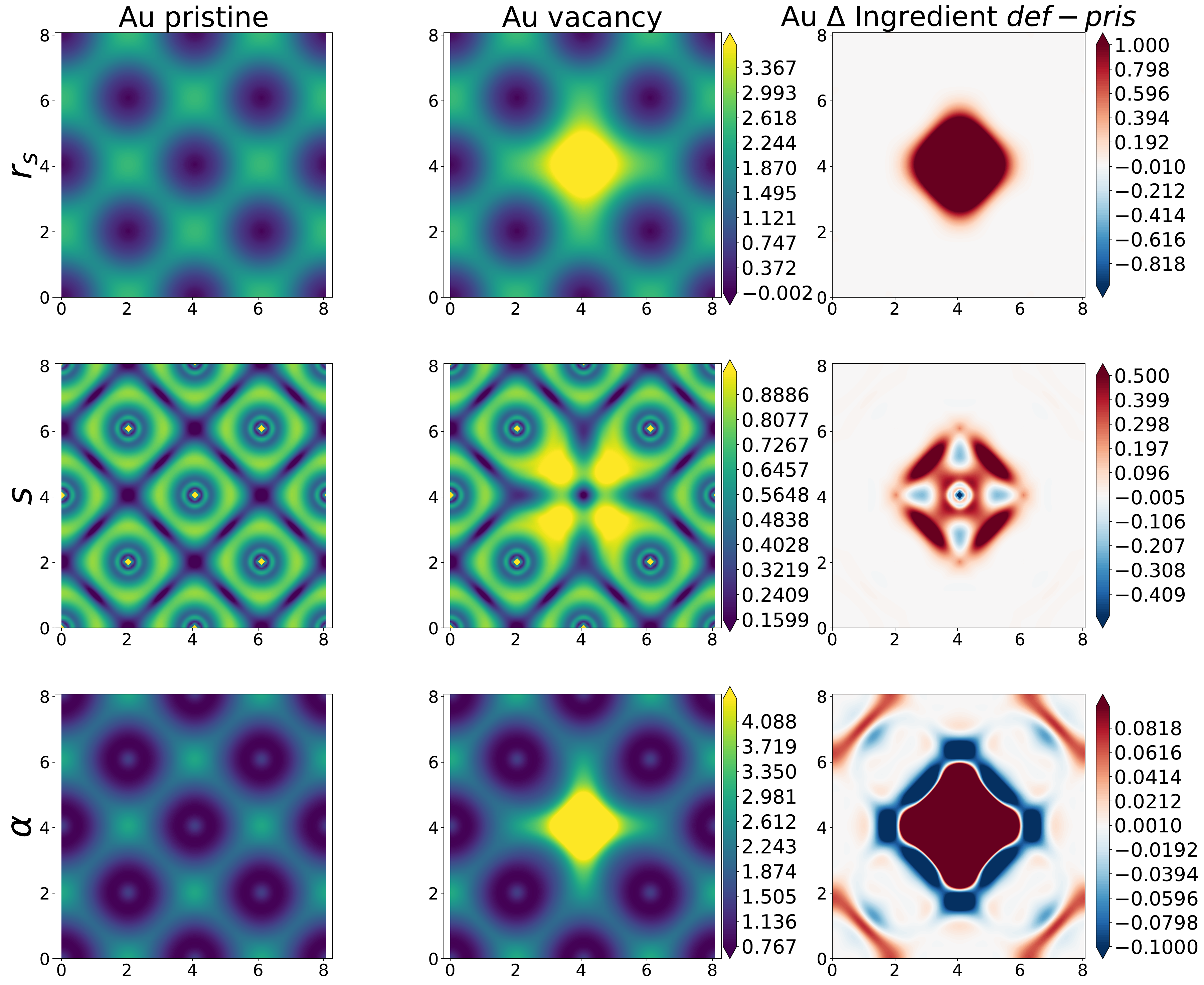}{Au}
\twodmetalfig{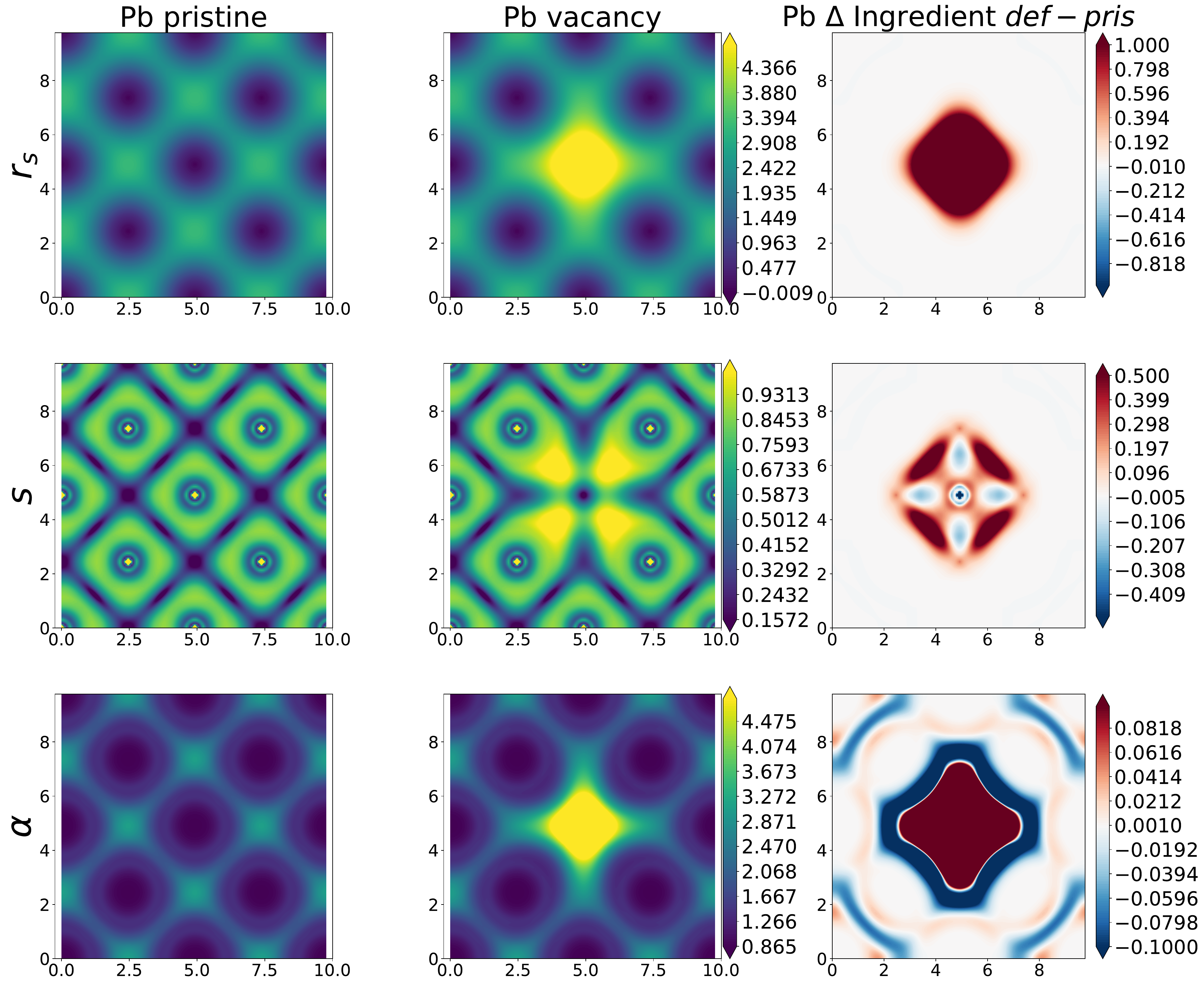}{Pb}

\clearpage
\subsection{Defect formation energies of Si interstitials at unrelaxed geometries}

\begin{figure}[h]
    \centering
\includegraphics[width=0.65\linewidth]{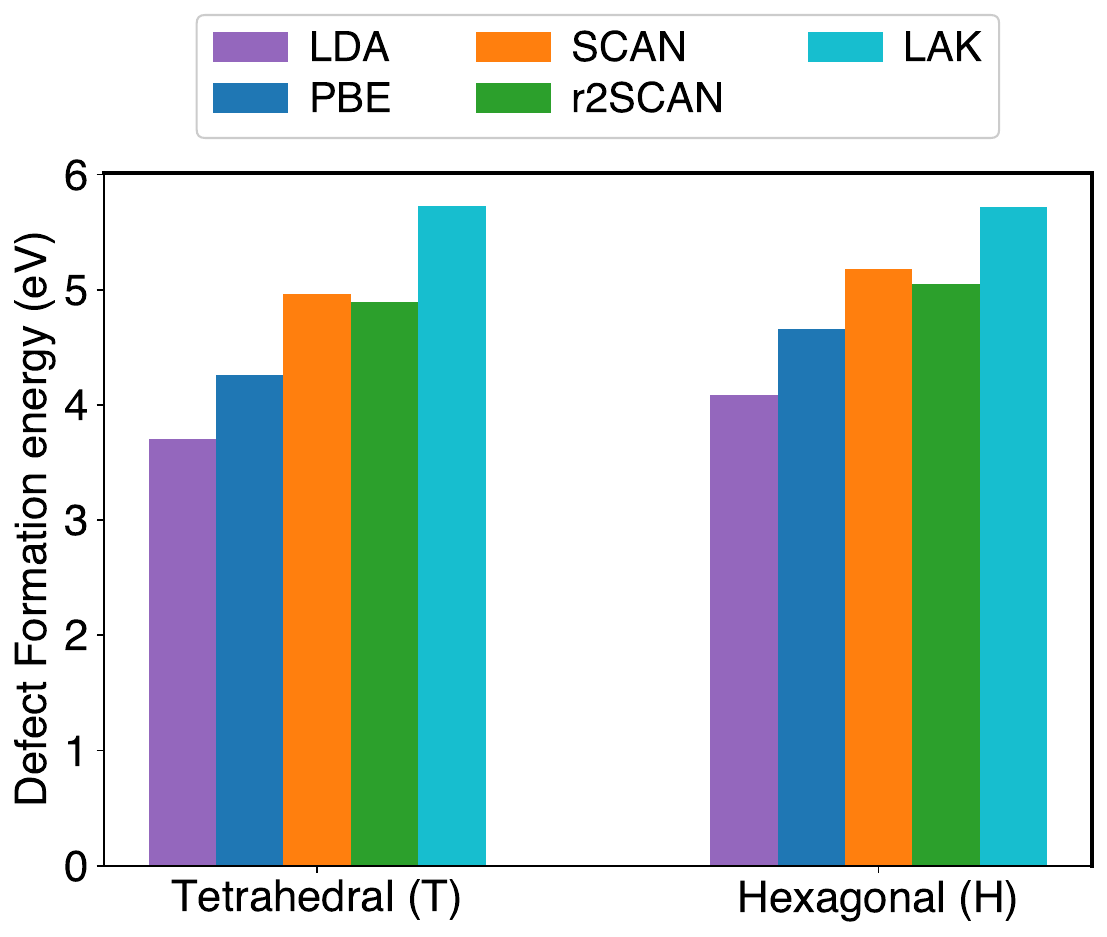}
 \caption{Formation energies of the T and H interstitial defects in Si-diamond for several DFAs calculated with the unrelaxed structures} 
 \label{si-interstitials_exp}
\end{figure}

Figure \ref{si-interstitials_exp} shows the defect formation energies for the T and H interstitials in Si using unrelaxed structures with the experimental lattice constants of the pristine system and fixed atomic positions. All DFAs follow the same trends as for the relaxed structures, compare Fig.~2 of the main text. As expected, the defect formation energies for the unrelaxed structures are larger than for the relaxed ones, since the unrelaxed structure is no experimental structure for the defective system. As a consequence, the effect of relaxation is larger in the defective system and the energy of the defective systems is reduced more under relaxation, reducing the defect formation energy.


\begin{table*}[htb!]
\centering
\caption{The interstitial formation energy in eV of Si diamond using the unrelaxed structures, i.e., with experimental lattice constants and no relaxation. Note that this does not correspond to an experimental structure for the defective system and is of limited practical relevance. Instead, this structure is considered to enable a transparent comparison of different DFAs in the ingredients analysis, which requires using the same structure for all DFAs.}
\begin{tabular}{c c c }
\hline\hline 
    & Tetrahedral (T) & Hexagonal (H)  \\
        LDA    & 3.70 & 4.09  \\
        PBE    & 4.26 & 4.66 \\
        SCAN   & 4.96 & 5.18  \\
        $r^2$SCAN & 4.89 & 5.05 \\
        LAK    & 5.73 & 5.72  \\
\hline
\hline
\end{tabular}
\label{tab:si_diamond_table_unrelaxed}
\end{table*}

\clearpage 
\bibliography{bibliography}